\documentclass[aps,prd,superscriptaddress,amsmath,groupedaddress,twocolumn,nofootinbib]{revtex4-1}

\usepackage[pdftex]{color}
\usepackage[dvipsnames]{xcolor}
\usepackage[sort&compress]{natbib}
\usepackage{color,hhline,psfrag,rotating,amssymb}
\usepackage[colorlinks=true,linkcolor=PineGreen,filecolor=PineGreen,urlcolor=PineGreen,citecolor=PineGreen,pdftex,plainpages=false]{hyperref}
\usepackage[hang,nooneline]{subfigure}
\usepackage{dcolumn}

\begin{document}
\title{Bottomonium precision tests from full lattice QCD: hyperfine splitting, $\Upsilon$ leptonic width and $b$ quark contribution to $e^+e^- \rightarrow$ hadrons. }

\author{D.~Hatton}
\email[]{daniel.hatton@glasgow.ac.uk}
\affiliation{SUPA, School of Physics and Astronomy, University of Glasgow, Glasgow, G12 8QQ, UK}
\author{C.~T.~H.~Davies}
\email[]{christine.davies@glasgow.ac.uk}
\affiliation{SUPA, School of Physics and Astronomy, University of Glasgow, Glasgow, G12 8QQ, UK}
\author{J.~Koponen}
\affiliation{Helmholtz Institute Mainz, Johannes Gutenberg-Universit\"{a}t Mainz, 55099 Mainz, Germany}
\author{G.~P.~Lepage}
\affiliation{Laboratory for Elementary-Particle Physics, Cornell University, Ithaca, New York 14853, USA}
\author{A.~T.~Lytle}
\affiliation{Department of Physics. University of Illinois, Urbana, IL 61801, USA}
\collaboration{HPQCD collaboration}
\homepage{http://www.physics.gla.ac.uk/HPQCD}
\noaffiliation

\date{\today}

\begin{abstract}
We calculate the mass difference between the $\Upsilon$ and $\eta_b$
and the $\Upsilon$ leptonic width 
from lattice QCD using the Highly Improved Staggered Quark 
formalism for the $b$ quark and including $u$, $d$,  
$s$ and $c$ quarks in the sea. 
We have results for lattices with lattice spacing as low as 0.03 fm 
and multiple heavy quark masses, enabling us to map out the heavy quark 
mass dependence and determine values at the $b$ quark mass. 
Our results are: $M_{\Upsilon} -M_{\eta_b} = 57.5(2.3)(1.0) \,\mathrm{MeV}$ (where 
the second uncertainty comes from neglect of quark-line disconnected correlation 
functions) and decay constants, $f_{\eta_b}=724(12)$ MeV and 
$f_{\Upsilon} =677.2(9.7)$ MeV, giving 
$\Gamma(\Upsilon \rightarrow e^+e^-) = 1.292(37)(3) \,\mathrm{keV}$. 
The hyperfine splitting and leptonic width are both in good 
agreement with experiment, and provide the most accurate lattice QCD results to date 
for these quantities by some margin. 
At the same time results for the time moments of the vector-vector correlation function 
can be compared to values for the $b$ quark contribution 
to $\sigma(e^+e^- \rightarrow \mathrm{hadrons})$ determined from experiment.  
Moments 4--10 provide a 2\% test of QCD and yield a $b$ quark contribution 
to the anomalous magnetic moment of the muon of 0.300(15)$\times 10^{-10}$. 
Our results, covering a range of heavy quark masses, may also be useful to constrain QCD-like
composite theories for beyond the Standard Model physics. 
\end{abstract}


\maketitle

\section{Introduction}
\label{sec:intro}

Weak decay matrix elements calculated in lattice QCD 
are critical to the flavour physics programme of 
over-determining the Cabibbo-Kobayashi-Maskawa (CKM) matrix 
to find signs of new physics. For that programme the weak decays of 
$b$ quarks are particularly important since they give access 
to the least well known CKM elements, $V_{ub}$ and $V_{cb}$. 
These CKM matrix elements can be determined either using exclusive
decay channels and lattice QCD form factors or inclusive decay channels 
and measured spectral shape functions. There continues to be 
some tension between 
the exclusive and inclusive determinations~\cite{Zyla:2020zbs}
that needs further improvements to both approaches to resolve. 
On the lattice QCD side this means developing improved approaches to 
$B$ meson weak decay matrix elements, such as~\cite{McLean:2019qcx, Harrison:2020gvo}, but 
also providing more stringent tests of lattice QCD results in 
$b$ physics to make sure that sources of systematic error are 
under full control. Here we provide three such tests using 
bottomonium correlation functions; the ground-state hyperfine splitting 
(the mass difference between the $\Upsilon$ and $\eta_b$), the $\Upsilon$ 
leptonic width 
and the $b$ quark contribution to $R(e^+e^- \rightarrow \mathrm{hadrons})$. 
The latter two, being electromagnetic processes, can be compared with 
experiment free from CKM uncertainties. 
We obtain the most accurate results to date for these quantities and are 
able to include the effect of the $b$ quark's electric charge in the calculation 
for the first time. 
 
We use the Highly Improved Staggered Quark (HISQ) discretisation 
of the quark action for these calculations. The HISQ action was 
developed in~\cite{hisqdef} to have small discretisation errors 
with the leading errors, quadratic in the lattice spacing, removed. 
This makes the action particularly good for heavier quarks when
discretisation errors appear as powers of the quark 
mass in lattice units, which can be relatively large. 
This action enabled the first accurate 
lattice calculations in charm physics~\cite{fdsorig,fdsupdate,na1,psipaper}. 
More recently it has been used to achieve sub-1\% accuracy in 
the charmonium hyperfine splitting, $J/\psi$ leptonic width, $m_c$ and $c$ 
quark vacuum polarisation contribution to the anomalous magnetic moment 
of the muon~\cite{Hatton:2020qhk}. The calculation used 
a range of lattice spacing values from 0.15 fm to 0.03 fm with $u$, $d$, 
$s$ and $c$ quarks in the sea and included 
the effect of the $c$ quark's electric charge~\cite{Hatton:2020qhk}. 

Here we will extend this latter calculation to bottomonium. 
Because the $b$ quark mass is much larger than that of $c$, 
we need fine lattices to reach the $b$ with a quark mass 
in lattice units, $am_b < 1$ and controlled discretisation errors. 
Our strategy, known as the heavy-HISQ approach~\cite{bshisq}, is to perform 
calculations for a range of masses between $c$ and $b$ on 
lattices with a range of fine lattice spacings. We can then map 
out the dependence on the heavy quark mass both 
of the quantity being calculated and its 
discretisation errors. 
This 
enables us to determine a physical result at the $b$ quark mass.

This approach has been very successful for decay constants and 
spectroscopy for heavy-light ($B$, $B_s$ and $B_c$) 
mesons~\cite{bshisq,McNeile:2012qf,Bazavov:2017lyh} and is now 
being used for the form factors for $B$ meson 
weak decays~\cite{McLean:2019qcx,Harrison:2020gvo}. 
Here we will apply this approach to the $\Upsilon$ for the 
first time. 

There are alternative nonrelativistic approaches 
that can be used at the $b$ quark 
mass on coarser lattices; see~\cite{Colquhoun:2014ica} for the 
determination of the $\Upsilon$ and $\Upsilon^{\prime}$ leptonic 
widths using lattice NonRelativistic QCD. 
The nonrelativistic expansion of the Hamiltonian and 
the currents that appear in matrix elements gives 
systematic uncertainties from missing higher-order 
relativistic corrections and from renormalisation of the lattice 
current to match the continuum current. 
These uncertainties hinder tests at high precision. 

In contrast the HISQ action is relativistic and the 
HISQ vector current can be matched accurately 
and fully nonperturbatively to that in continuum QCD~\cite{Hatton:2019gha}.
As we will show below, this enables us to improve the lattice 
QCD accuracy on the bottomonium hyperfine splitting to better 
than 5\% and to achieve percent-level precision on the $\Upsilon$ 
and $\eta_b$ decay constants and on moments that parameterise the 
$b$ quark contribution to $R(e^+e^- \rightarrow \mathrm{hadrons})$. 

Our results cover the range of heavy quark masses from $c$ to $b$ 
and we will give results for decay constant to mass ratios over this 
range. These could be useful both for tuning of phenomenological models 
of QCD but also as constraints on QCD-like composite theories for beyond 
the Standard Model physics. 

The paper is organised as follows. In the next Section we give 
details of the lattice calculation we perform. This includes 
a general description of the fits that we use to determine the heavy 
mass dependence of the quantities calculated. We then
present results for the bottomonium hyperfine splitting in 
Section~\ref{sec:hyp}, the decay constants for both the 
$\eta_b$ and $\Upsilon$ in Section~\ref{sec:decconsts} and the 
time-moments of the vector current-current correlators in 
Section~\ref{sec:time-moms}. 
Each section includes a description of the calculation and then 
a discussion subsection with comparison to experiment and 
previous lattice QCD results. Section~\ref{sec:decconsts} on decay constants 
also includes plots of decay constant to mass ratios and the ratio of vector to pseudoscalar 
decay constants over the quark mass range 
from $c$ to $b$.  
We then give our conclusions in Section~\ref{sec:conclusions}.

\section{Lattice calculation}
\label{sec:latt}

We use ensembles of lattice gluon field 
configurations provided by the MILC collaboration~\cite{Bazavov:2012xda} 
at values of the lattice spacing, $a \approx$ 0.09 fm, 
0.06 fm, 0.045 fm and 0.03 fm. 
The configurations are generated with an $\alpha_s a^2$-improved discretisation of the gluon 
action~\cite{Hart:2008sq} and include the effect of $u$, $d$, $s$ and $c$ 
quarks in the sea with the HISQ formalism~\cite{hisqdef}. 
The $u$ and $d$ masses are taken to be the 
same and we denote this mass $m_l$. 
For most of the ensembles we have unphysically heavy $u/d$ quarks with 
$m_{l}/m_s \approx$ 0.2 
but we employ two ensembles with physical values of $m_l$
and lattice spacing values ~0.09 fm and ~0.06 fm.
We expect sea quark mass effects to be small 
for the $\Upsilon$ because it has no valence light
quarks. However, an analysis of such effects is needed for accurate results. 

Table~\ref{tab:params} lists the 
parameters of the ensembles that we use. 
The lattice spacing is determined 
in terms of the Wilson flow parameter $w_0$~\cite{Borsanyi:2012zs}.
On these ensembles we calculate quark propagators 
from random wall sources using the HISQ action
and
with a variety of masses, $m_h$, from that of the $c$ quark upwards.
The valence heavy quark masses that we use on each ensemble are 
listed in Table~\ref{tab:massres}.
The value of $\epsilon_{\mathrm{Naik}}$ used in the coefficient of the Naik 
term in the HISQ action~\cite{hisqdef} is taken as the tree-level function of 
the quark mass given in~\cite{fdsupdate}. 

We combine the quark propagators into 
(connected) meson correlation functions for both pseudoscalar ($\eta_h$) 
and vector ($\phi_h$) mesons, using the local $\gamma_5$ and $\gamma_i$ operators 
converted to appropriate form for staggered quarks~\cite{psipaper,Hatton:2020qhk}. Note that we do not include quark-line disconnected corrrelation functions that take account of the heavy quark/antiquark annihilation to gluons.
We expect the effect of the disconnected correlation functions 
to be very small in the heavyonium system. 
In~\cite{Hatton:2020qhk} our result for the 
mass difference between $J/\psi$ and $\eta_c$ mesons was accurate 
enough, for the first time, to see a difference 
with experiment of 7.3(1.2) MeV. 
We concluded that this was the effect of the missing 
disconnected correlation function on the $\eta_c$ mass. Here we will test 
for a similar effect on the $\eta_b$. 

On the coarsest two ensembles we use 16 time-sources 
on each gluon field configuration for high statistics;
we take 8 time-sources on the other ensembles. We use at least 100 
configurations on each ensemble for a good statistical sample.
In generating the very fine lattice (set 6 in Table~\ref{tab:params}) 
a slow evolution in Monte 
Carlo time of the topological charge was observed~\cite{Bazavov:2012xda}, so that the ensemble 
does not explore many topological sectors. 
However, 
it has been shown that the impact of this on calculations for 
heavy mesons is negligible~\cite{Bernard:2017npd}. 

\begin{table}
\caption{Sets of MILC configurations~\cite{Bazavov:2012xda} 
used here with 
HISQ sea quark masses, $m_l^{\mathrm{sea}}$ ($l=u/d$), 
$m_s^{\mathrm{sea}}$ and $m_c^{\mathrm{sea}}$ given in lattice units.  
The lattice spacing is given in units
of $w_0$~\cite{Borsanyi:2012zs}; the physical value of 
$w_0$ was determined to be 0.1715(9) fm from $f_{\pi}$~\cite{fkpi}. 
Sets 1 and 2 are `fine' ($a \approx 0.09$ fm), sets 3 and 4 are
`superfine' ($a \approx 0.06$ fm), set 5 `ultrafine' ($a \approx 0.045$ fm) 
and set 6 `exafine' ($a \approx 0.03$ fm).  
The final two columns give the extent of the lattice in each spatial 
direction ($L_s$) and time ($L_t$). 
}
\begin{tabular}{lllllllll}
\hline
\hline
Set & label & $w_0/a$ & $am_{l}^{\mathrm{sea}}$ & $am_{s}^{\mathrm{sea}}$ & $am_{c}^{\mathrm{sea}}$ & $L_s$ & $L_t$ \\
\hline
1 & f-5 & 1.9006(20) & 0.0074 & 0.037 & 0.440 & 32 & 96  \\
2 & f-phys & 1.9518(7) & 0.00120 & 0.0363 & 0.432 & 64 & 96 \\
\hline
3 & sf-5 & 2.8960(60) & 0.00480 & 0.0240 & 0.286 & 48  & 144 \\
4 & sf-phys & 3.0170(23) & 0.0008 & 0.022 & 0.260 & 96 & 192 \\
\hline
5 & uf-5 & 3.892(12) & 0.00316 & 0.0158 & 0.188 & 64 & 192 \\
\hline
6 & ef-5 & 5.243(16) & 0.00223 & 0.01115 & 0.1316 & 96 & 288 \\
\hline
\hline
\end{tabular}
\label{tab:params}
\end{table}

\begin{table*}
\caption{ Results in lattice units for the masses of the ground-state 
pseudoscalar meson, $\eta_h$, and ground-state vector 
meson, $\phi_h$, for valence heavy quark masses in 
lattice units listed in column 2, for the ensembles listed in column 1.   
Results come from simultaneous fits to all heavy quark masses on a given 
ensemble, except for the cases marked with an asterisk~\cite{Hatton:2020lnm}. These used different 
random numbers for the sources and so are not correlated with the other results for that ensemble. 
Column 5 gives the mass difference in lattice units between the $\phi_h$ 
and $\eta_h$, column 6 the $\eta_h$ decay constant and column 7 the 
raw (unrenormalised) $\phi_h$ decay constant. The required $Z_V$ factors 
are taken from~\cite{Hatton:2019gha}.
 }
\begin{tabular}{lllllll}
\hline
\hline
Set & $am_h$ & $aM_{\eta_h}$ & $aM_{\phi_h}$ & $a\Delta M_{\mathrm{hyp}}$  & $af_{\eta_h}$ & $af_{\phi_h}/Z_V$ \\
\hline
1 & 0.6 & 1.675554(47) & 1.717437(70) & 0.041882(84) & 0.208641(60) & 0.21865(11) \\
  & 0.8 & 2.064088(40) & 2.101542(57) & 0.037454(70) & 0.249695(64) & 0.25711(10) \\
\hline
2 & 0.6 & 1.674264(13) & 1.715453(32) & 0.041189(35) & 0.207535(22) & 0.21690(10) \\
 & 0.8 & 2.063015(11) & 2.099940(26) & 0.036925(29) & 0.248493(21) & 0.255249(96) \\
 & 0.866$^*$ & 2.185464(53) & 2.221789(38) & 0.036325(65) & 0.264483(61) & 0.27011(12) \\
\hline
3 & 0.274 & 0.896664(33) & 0.929876(86) & 0.033212(92) & 0.117554(37) & 0.12339(15) \\
 & 0.4 & 1.175559(29) & 1.202336(85) & 0.026778(90) & 0.135692(39) & 0.13916(21) \\
 & 0.5 & 1.387459(27) & 1.411113(72) & 0.023654(77) & 0.148936(40) & 0.15104(20) \\
 & 0.548$^*$ & 1.487111(36) & 1.509697(54) & 0.022586(65) & 0.155563(68) & 0.157200(87) \\
 & 0.6 & 1.593089(25) & 1.614626(63) & 0.021537(68) & 0.162314(41) & 0.16318(19) \\
 & 0.7 & 1.793118(23) & 1.813249(57) & 0.020131(61) & 0.176638(42) & 0.17617(19) \\
 & 0.8 & 1.987504(22) & 2.006783(52) & 0.019279(56) & 0.192680(44) & 0.19061(19) \\
\hline
4 & 0.260 & 0.862671(27) & 0.895702(52) & 0.033030(58) & 0.114147(34) & 0.11969(10) \\
 & 0.4 & 1.173904(23) & 1.199806(36) & 0.025903(43) & 0.134475(37) & 0.137266(83) \\
 & 0.6 & 1.591669(19) & 1.612586(27) & 0.020917(34) & 0.161035(39) & 0.161236(77) \\
 & 0.8 & 1.986246(17) & 2.005047(24) & 0.018801(29) & 0.191297(41) & 0.188634(82) \\
\hline
5 & 0.194 & 0.666821(41) & 0.692026(59) & 0.025205(72) & 0.087774(42) & 0.091442(91) \\
 & 0.4 & 1.130722(31) & 1.147617(40) & 0.016895(51) & 0.114953(46) & 0.114918(72) \\
 & 0.6 & 1.549098(26) & 1.562884(32) & 0.013786(41) & 0.137487(54) & 0.135412(69) \\
 & 0.8 & 1.945787(23) & 1.958252(27) & 0.012465(35) & 0.162850(58) & 0.158238(72) \\
 & 0.9 & 2.135642(21) & 2.147903(25) & 0.012261(33) & 0.178229(58) & 0.171745(74) \\
\hline
6 & 0.138 & 0.496969(42) & 0.516149(61) & 0.019180(74) & 0.065916(59) & 0.06841(10) \\
 & 0.45 & 1.201328(29) & 1.211601(28) & 0.010273(40) & 0.102989(81) & 0.100572(70) \\
 & 0.55 & 1.410659(27) & 1.420048(24) & 0.009389(36) & 0.112668(82) & 0.109506(67) \\
 & 0.65 & 1.614877(24) & 1.623684(21) & 0.008807(32) & 0.122639(81) & 0.118680(64) \\

\hline \hline
\end{tabular}
\label{tab:massres}
\end{table*}

We fit the correlation functions from each ensemble 
using a multi-exponential constrained fit~\cite{gplbayes} and following
the method in~\cite{Hatton:2020qhk}. 
The fit form used for the pseudoscalar correlators as a function of 
$t$, the time separation between source and sink, is 
\begin{equation} \label{eq:ps-corr-fit}
C_P(t) = \sum_i A_i^P f(E_i^P,t) ,
\end{equation}
and the vector fit form is 
\begin{equation}
\label{eq:v-corr-fit}
C_V(t) = \sum_i \left( A_i^V f(E_i^V,t) - (-1)^t A_i^{V,o} f(E_i^{V,o},t) \right) .
\end{equation}
Here
\begin{equation}
f(E,t) = e^{-Et} + e^{-E(L_t-t)} .
\end{equation}
The term that oscillates in time in the vector case results from the use of 
staggered quarks. $E_0$ is the mass of the lowest lying state 
(either pseudoscalar or vector) and $A_0$ is related to the meson 
decay constants. The ground-state pseudoscalar meson we will denote as 
$\eta_h$ and the vector as $\phi_h$. We fit the correlation functions for 
all masses on a given ensemble simultaneously (with two exceptions, see 
Table~\ref{tab:massres}). This means that the correlations between results 
for different masses are carried through the rest of the calculation. 
The correlations between the 
$\phi_h$ and $\eta_h$ correlators are safely neglected as 
the uncertainty in the $\phi_h$ results dominates that for the $\eta_h$.
Results for the ground-state mesons are listed in Table~\ref{tab:massres}. 

\begin{table*}
\caption{Quenched QED corrections, for quark electric charge $e/3$, 
to a subset of the results of Table~\ref{tab:massres} presented 
as the ratio, $R^0$, of the value in QCD+QED to that in pure QCD at fixed valence quark mass in lattice units.}
\begin{tabular}{lllllll}
\hline \hline
Set & $am_h$ & $R^0_{\mathrm{QED}}[aM_{\eta_h}]$ & $R^0_{\mathrm{QED}}[aM_{\phi_h}]$ & $R^0_{\mathrm{QED}}[a\Delta M_{\mathrm{hyp}}]$  & $R^0_{\mathrm{QED}}[af_{\eta_h}]$ & $R^0_{\mathrm{QED}}[af_{\phi_h}/Z_V]$ \\
\hline
2 & 0.866 & 1.0002170(45) & 1.0002637(14) & 1.00307(25) & 1.001255(68) & 1.001343(62) \\
\hline
3 & 0.274 & 1.0003937(42) & 1.0004468(28) & 1.001875(83) & 1.00078(12) & 1.000708(83) \\
 & 0.548 & 1.0003306(12) & 1.00036876(24) & 1.002882(80) & 1.001127(35) & 1.0009805(75) \\
\hline \hline
\end{tabular}
\label{tab:qedres}
\end{table*}

We also have a limited amount of data which includes the 
effects of quenched QED (electrically charged 
valence quarks, but not sea quarks). This allows us to 
assess the impact of QED and appropriately account for 
it in our error budgets. As in~\cite{Hatton:2020qhk} we 
use photon fields in Feynman gauge in the $\mathrm{QED}_L$ 
formalism~\cite{Hayakawa:2008an}.
Our quenched QED calculations~\cite{Hatton:2020qhk, Hatton:2020lnm} used a valence quark 
electric charge of 2/3$e$ (i.e. the charge on a $c$ quark), where $e$ is the magnitude 
of the charge on an electron. 
We can use these results to determine
the electromagnetic correction for the $b$ electric charge of $-(1/3)e$. 
Given the smallness of $\alpha_{\mathrm{QED}}$ 
we take QED corrections to be linear in the quark charge squared, 
$Q^2$, and simply rescale the effect of QED by a factor of 1/4 from that for $Q=(2/3)e$. 
Results are given in Table~\ref{tab:qedres} 
in the form of the ratio, $R^0$, of results in QCD+QED to those in pure QCD
at a fixed value of the valence quark mass in lattice units. 
We see there that the impact of QED is tiny but visible. 
We showed in~\cite{Hatton:2020qhk} that QED finite-volume effects 
were negligible for the electrically neutral charmonium mesons. 
This will continue to be true for the heavier mesons that we study 
here and so we ignore such effects.  

As we showed in~\cite{Hatton:2020qhk}, fixing the lattice spacing from 
$w_0$ and $f_{\pi}$, as we have done, means that QED corrections to the 
lattice spacing should be at the sub-0.1\% level 
(coming from QED effects in the quark sea).  We can therefore compare 
QCD plus quenched QED to pure QCD using the same value of the 
lattice spacing (i.e. that from Table~\ref{tab:params}).  
QED affects the tuning of the lattice quark masses, however. 
We use the simple and natural scheme of tuning the $b$ quark mass in 
both the QCD+QED and pure QCD cases so that the $\Upsilon$ 
mass has the physical value. We can then use our results to determine the 
impact of QED on the quantities we study, taking the renormalisation 
of the quark mass into account. We will give that information, 
after fitting, 
as the renormalised ratio $R$ which is the ratio of the QCD+QED 
result to that in pure QCD when both theories have a $b$ quark mass 
separately tuned so that the $\Upsilon$ mass takes the experimental 
value in both cases. 

For each quantity that we study we must 
fit our results, in
physical units, as a function of heavy quark mass and 
lattice spacing to determine results in the continuum 
limit at the physical $b$ quark mass. 
We will use the $\phi_h$ mass as a physical proxy for the 
heavy quark mass and then the physical point is defined by 
the $\phi_h$ mass becoming equal to that of the $\Upsilon$. 

In previous studies of heavy meson masses and decay constants using the heavy-HISQ method 
the HPQCD collaboration have used fit forms that capture the 
heavy mass dependence as a polynomial in the inverse $H_s$ or $\eta_h$ 
mass~\cite{bshisq, McNeile:2012qf}. 
In the case of heavy-light mesons this form is justified 
by the heavy quark effective theory (HQET) expansion. 
In the 
case of heavyonium the HQET expansion is not valid but the same 
form may still be expected to work as a Taylor expansion over a finite 
region in $m_h$. 
Here, however, we choose to use a fit form 
that is more agnostic with regards to the dependence on the heavy quark mass. 
We achieve this by using cubic splines between specified knot positions\footnote{ 
We use splines that are monotonic between knots (Steffen splines~\cite{Steffen:1990}). }.
We do not expect to need many knots because the 
quantities we study here 
should be smooth monotonic functions 
of $M_{\phi_h}$ in the continuum limit at physical sea quark masses. 
The fit function for our lattice results also needs to include dependence 
on the lattice spacing and the mistuning of sea quark masses. 
Both of these effects can also depend on 
the heavy quark mass ($M_{\phi_h}$) through smooth monotonic functions 
and so we also include cubic splines in 
their description. 

We use fits of the following form for the pure QCD part of our fit:
\begin{eqnarray}
\label{eq:spline-fit}
F(a,M_{\phi_h}) [\mathrm{QCD}] &=& A \bigl[ F_0(M_{\phi_h}) + G_0(1/M_{\phi_h}) \\ &+& \sum_{i=1}^3  G_1^{(i)}(M_{\phi_h}) (am_h)^{2i} \nonumber \\
&+& \sum_{j=1}^3  G_2^{(j)}(M_{\phi_h}) (a\Lambda)^{2j} \nonumber \\ 
&+&   G_3(M_{\phi_h}) (a\Lambda)^{2}(am_h)^2 \nonumber \\ 
&+& \sum_{k=0}^2 G_4^{(k)}(M_{\phi_h})(am_h)^{2k} \delta_l \nonumber \\ 
&+&  G_5(M_{\phi_h}) \delta_c \bigr] \nonumber 
\end{eqnarray}
$F(a,M_{\phi_h})$ are the lattice QCD results in physical 
units of GeV for the hyperfine splitting and decay constants and 
$\mathrm{GeV}^{-1}$ for the time-moments. 
$A$ is then a dimensionful number of a size commensurate with the size 
of the quantity being fitted (this will be given in each section) so 
that the rest of the fit in square brackets is dimensionless. 
$F_0$ is a dimensionless function of $M_{\phi_h}$ that differs between 
the different quantities we examine (and will also be given in each section). 
It is a very simple function of $M_{\phi_h}$
that captures the general 
trend in the mass dependence, on top of which the corrections
modelled by splines are relatively minor. 
$G_n$ denotes a cubic spline. All of the splines have 
different parameters but keep the same positions for the knots. 
The splines are functions of $M_{\phi_h}$, where $M_{\phi_h}$ is in GeV, 
except for 
the first spline, $G_0$, which provides the physical corrections to $F_0$. 
We found that a spline in $1/M_{\phi_h}$ rather than $M_{\phi_h}$ gave a better 
$\chi^2$ for quantities like the hyperfine splitting which fall as $M_{\phi_h}$ grows, 
approximately as the inverse. 
We allow for two kinds of discretisation effects, those that are set by the 
heavy quark mass $am_h$ and those that are independent of the heavy quark mass 
and instead are set by a fixed scale that we call $\Lambda$. We take $\Lambda$ to be 0.5 GeV. 
The $G_1^{(i)}$ splines 
allow for heavy mass dependence in the $am_h$ discretisation 
effects and the $G_2^{(j)}$ splines allow for heavy mass dependence in the 
$a\Lambda$ discretisation effects. 
We also include a mixed term in $(am_h)^2(a\Lambda)^2$ with spline 
$G_3$, although this has little impact on the fits. 

The last two lines of the fit form in Eq.~(\ref{eq:spline-fit}) allow 
for sea quark mass mistunings. The light quark mass mistuning 
parameter $\delta_l$ is defined as:
\begin{equation}
\delta_l = \frac{2m_l^{\mathrm{sea}} + m_s^{\mathrm{sea}} - 2m_l^{\mathrm{phys}} - m_s^{\mathrm{phys}}}{10m_s^{\mathrm{phys}}} .
\end{equation}
$m_s^{\mathrm{phys}}$ and $m_l^{\mathrm{phys}}$ are taken to 
be the same as those used in~\cite{Hatton:2020qhk}. The charm 
quark mass mistuning parameter is defined similarly:
\begin{equation}
\delta_c = \frac{m_c^{\mathrm{sea}} - m_c^{\mathrm{phys}}}{10 m_c^{\mathrm{phys}}} .
\end{equation}
We allow for discretisation effects and 
heavy mass dependence set by splines $G_4^{(k)}$ within the light 
sea quark mass mistuning term. 
We allow for heavy mass dependence in the 
sea charm mass mistuning term through the spline $G_5$. 

For data that includes quenched QED effects we add extra terms 
to the fit function in Eq.~\eqref{eq:spline-fit}. The full fit 
takes the form:
\begin{eqnarray} \label{eq:spline-fit-qed}
&&F(a,M_{\phi_h}) [\mathrm{QCD}+\mathrm{QED}] = F(a,M_{\phi_h}) [\mathrm{QCD}] + \\
&&A\alpha_{\mathrm{QED}} Q^2 \left[ \hat{G}_1(1/M_{\phi_h}) + c_{\mathrm{QED},am_h} \hat{G}_2(M_{\phi_h}) (am_h)^2 \right]. \nonumber
\end{eqnarray}
$\hat{G}_1$ and $\hat{G}_2$ are additional spline functions. 

For all the fits considered here we use knots placed at $\{2.5,4.9,10\}$ 
GeV, taking knots at the beginning and end of the fit range and one in the middle. 
This is the optimal number of knots according to the Bayes Factor. 
We have checked that varying the sums over discretisation effects so that the 
number of terms included changes by $\pm 1$ has no significant effect. 
Because statistical errors are so small here (see Table~\ref{tab:massres}) we employ an 
SVD (singular value decomposition) cut~\cite{Dowdall:2019bea} in the fit 
to account for tiny effects that are too small to be modelled by our fit form of 
Eq.~\ref{eq:spline-fit}. 
The SVD cut value times the maximum eigenvalue of the covariance matrix sets 
a minimum value for the eigenvalues. Any eigenvalues of the 
covariance matrix below this minimum are then replaced with the minimum value. 
This is a conservative move which increases our uncertainties, 
as can be seen from the impact of the SVD cut in our error 
budgets.

The prior information given to the fit are central values and widths for 
the values of the coefficients, $c_F$, in $F_0$ and for the values of the spline functions 
at each knot. We use priors of $0\pm 1$ for all of these. 
We use the \texttt{lsqfit} python module~\cite{peter_lepage_2020_4037174} to do the fits, implementing the splines with 
the \texttt{gvar} module~\cite{peter_lepage_2020_4290884}.

To obtain our final results, 
the fit function is evaluated at lattice spacing equal to zero, sea quark masses 
tuned to their physical values and $M_{\phi_h}$ equal to $M_{\Upsilon}$. 
This is taken from experiment as 9.4603 GeV~\cite{Zyla:2020zbs} with negligible uncertainty.

\begin{figure}
\includegraphics[width=0.9\hsize]{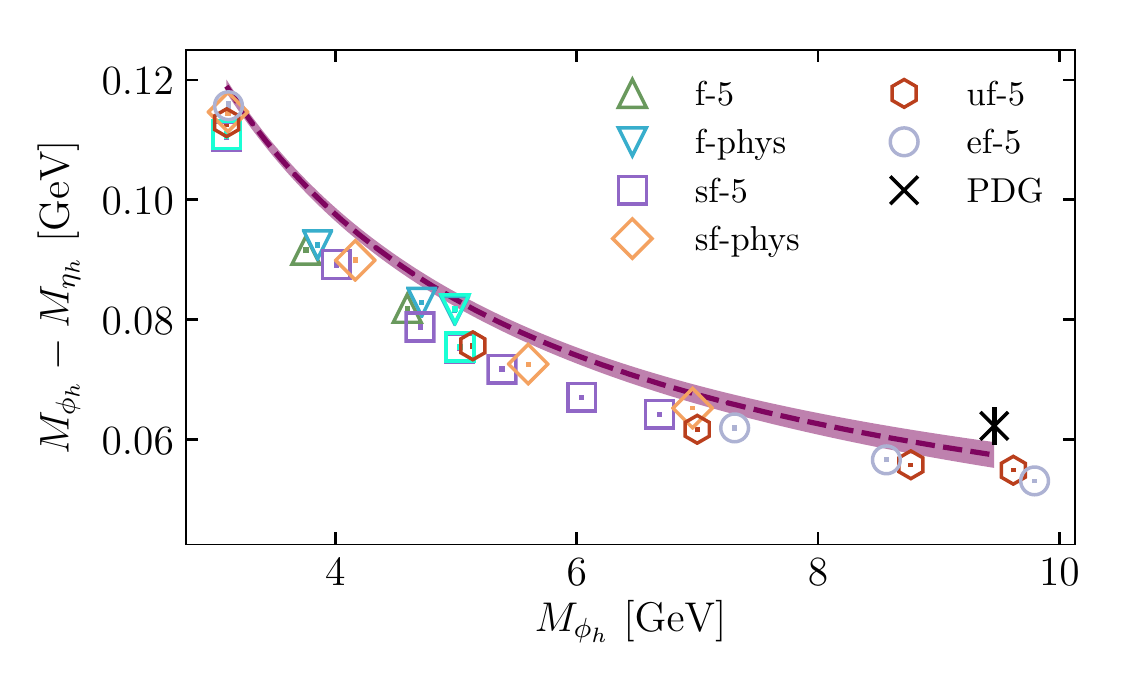}
\caption{The heavyonium hyperfine splitting 
as a function of the vector heavyonium mass, $M_{\phi_h}$. 
The points show our lattice results 
 from Tables~\ref{tab:massres} and~\ref{tab:qedres}, with 
different symbols denoting different ensembles as in the legend.
The errors are dominated by uncertainties from $a$ 
that are
correlated between the points. 
QCD+QED points are shown 
in cyan. They are not distinguishable from their pure QCD 
counterparts, but are visible by being shown on top of these values. 
Cubic splines are used to fit the heavy mass 
dependence, as described in the text. The fit evaluated at the physical point 
and zero lattice spacing is given by purple dashed line with error band. 
The experimental average value for the hyperfine splitting~\cite{Zyla:2020zbs} 
is plotted as the black cross
at the physical $\Upsilon$ mass. 
}
\label{fig:hyp}
\end{figure}

\section{Hyperfine splitting}
\label{sec:hyp}

The hyperfine splitting, $\Delta M_{\mathrm{hyp}}$, is the difference
in mass between the ground-state $\phi_h$ and $\eta_h$ mesons. 
The values for the hyperfine splitting on each ensemble 
for a variety of heavy quark masses are given in lattice units 
in column 5 of Table~\ref{tab:massres}. The 
separate $\eta_h$ and $\phi_h$ masses are given in 
lattice units in columns 3 and 4. The impact of quenched QED at 
fixed valence quark mass is given in Table~\ref{tab:qedres}. 
The effect of QED is similar to that for charmonium~\cite{Hatton:2020qhk} 
but reduced because of the smaller electric charge of the $b$ quark. 
The direct effect of QED on the hyperfine splitting is to increase
it, through a QED hyperfine effect which has the same sign 
as the QCD hyperfine effect. QED also increases the meson masses, however, and 
this requires a retuning of the bare quark masses downwards to match the 
same meson mass. This then has an indirect effect, increasing the hyperfine 
splitting by a very small amount.  

Our lattice results are plotted in Fig~\ref{fig:hyp} as a function of 
the vector heavyonium mass, $M_{\phi_h}$. 
The points include both pure QCD and QCD+QED values but the 
QCD+QED values are indistinguishable from pure QCD on this scale. 

To fit our results for the hyperfine splitting 
using Eq.~(\ref{eq:spline-fit})
we take $A =$ 0.1 GeV and the simple form for $F_0$, $F_0 = c_F^{(0)} + c_F^{(1)}(3\,\mathrm{GeV})/M_{\phi_h}$. 
The coefficients $c_F^{(0)}$ and $c_F^{(1)}$ have prior values $0(1)$. 
Note that we multiply the QED correction term in the fit (Eq.~(\ref{eq:spline-fit-qed})) 
by a factor of 2 because of the size of the QED corrections that we see 
in the results (Table~\ref{tab:qedres} and~\cite{Hatton:2020lnm}). 
That these prior widths are very conservative can be judged from 
the values and variation across Fig.~\ref{fig:hyp}. 
Evaluating the fit result at zero lattice spacing, tuned 
quark masses and with $M_{\phi_h}$ equal to the $\Upsilon$ 
mass, we obtain
the physical result for the bottomonium hyperfine splitting using 
connected correlation functions of:
\begin{equation}
M_{\Upsilon} - M_{\eta_b}\,\, \mathrm{(connected)} = 57.5(2.3)\ \mathrm{MeV} .
\label{eq:hypgive}
\end{equation}
This is the QCD+QED value. For the ratio of the QCD+QED value 
to the pure QCD result, we obtain:
\begin{equation}
R_{\mathrm{QED}} [\Delta M_{\mathrm{hyp}}] = 1.0001(26) .
\end{equation}
Note that this is the `renormalised' ratio with the $b$ quark mass 
tuned from the $\Upsilon$ in both QCD+QED and QCD. We see no significant 
impact of quenched QED at the 0.2\% level. 
The fit has a $\chi^2/\mathrm{dof}$ of 0.73 using an SVD cut of 
$5 \times 10^{-3}$.

Fig.~\ref{fig:hyp} shows our fit curve as a function of $M_{\phi_h}$ in 
the continuum limit for tuned sea quark masses.  
This gives useful physical insight into how the hyperfine splitting falls 
as the quark mass increases. At the high mass end of the plot we mark with a 
black cross the experimental average value~\cite{Zyla:2020zbs} for 
the bottomonium system. We will discuss the comparison to experiment further 
in Section~\ref{sec:discusshyp}. We note that at the lower mass end of 
the curve we have results for charmonium. Our fit here does not 
include all of the charmonium results that went into~\cite{Hatton:2020qhk}
but gives a value for the charmonium hyperfine splitting 
that is consistent (within 1$\sigma$) with~\cite{Hatton:2020qhk} 
for the pure QCD case.  
The QED+QCD result here is too small at the charmonium end of the fit 
curve because the QED is being included with quark 
charge 1/3$e$ rather than 
the correct charm quark charge of 2/3$e$. 

We will discuss in Section~\ref{sec:discusshyp} what the impact 
of quark-line disconnected (but gluon-connected) correlation 
functions could be on 
the bottomonium hyperfine splitting. For our charmonium calculation 
of~\cite{Hatton:2020qhk} we included an estimate of the QED quark-line 
disconnected 
contribution to the hyperfine splitting coming 
from $c\overline{c}$ annihilation to 
a single photon, which then converts back to $c\overline{c}$.   
The contribution of this to the charmonium hyperfine splitting 
is 0.7 MeV, which was a little more than half the uncertainty in our 
result. The equivalent contribution for the $\Upsilon$ here is much 
smaller, at 0.17 MeV, because of the smaller electric charge of 
the $b$ quark. At a size of one tenth of the uncertainty in our 
result in Eq.~\eqref{eq:hypgive}, this would then have 
negligible impact 
and we do not include it. 

A complete error budget for the bottomonium hyperfine splitting 
is given in Table~\ref{tab:errorbudget}. 
Statistical uncertainties are divided 
between those arising from our 2-point fits and those coming 
from the lattice spacing determination, both correlated between ensembles ($w_0$) 
and uncorrelated ($w_0/a$). The uncertainty from the 2-point fits 
is further divided in two. As already mentioned, the use of an SVD cut 
is conservative and increases the uncertainty in the fit output. 
We can calculate the contribution to an error budget of both the data 
with and without the SVD cut applied to its correlation matrix. In the 
error budgets of Table~\ref{tab:errorbudget} we give the contribution 
from the data with the original correlation matrix under the heading 
``statistics". The additional contribution from the SVD cut is then defined as the 
square root of the difference of the squared contributions from the data 
with and without an SVD cut applied. The contributions from various parts of the heavy mass 
dependence in Eqs.~\eqref{eq:spline-fit} and~\eqref{eq:spline-fit-qed} 
are shown individually, labelled by the set of spline functions for that contribution.

The fit parameters required to reproduce the physical curve of the hyperfine 
splitting as a function of $M_{\phi_h}$ plotted in Fig.~\ref{fig:hyp}  
are given in Table~\ref{tab:hyperfine-fit} of Appendix~\ref{sec:appendix}. 

\begin{table}
\caption{Error budget for the hyperfine splitting and decay constants as a percentage of the final 
answer.}
\begin{tabular}{lccc}
\hline
\hline
 &  \hspace{-1em}$M_{\Upsilon}-M_{\eta_b}$ & $f_{\Upsilon}$ \quad & \quad $f_{\eta_b}$ \\
\hline
statistics & 2.40 & 0.77 & 0.38 \\
SVD cut & 1.48 & 0.44 & 0.67 \\
$w_0$ & 0.55 & 0.61 & 0.59 \\
$w_0/a$ & 0.66 & 0.23 & 0.18 \\
$Z_V$ & - & 0.29 & - \\
$F_0$ & 0.03 & 0.01 & 0.00\\
$G_0$ & 0.05 & 0.02 & 0.01\\
$G_1$ & 1.14 & 0.17 & 0.18\\
$G_2$ & 0.48 & 0.24 & 0.31\\
$G_3$ & 0.42 & 0.28 & 0.45\\
$G_4$ & 1.45 & 0.73 & 0.98\\
$G_5$ & 1.08 & 0.29 & 0.27\\
$\hat{G}_1$ & 0.29 & 0.07 & 0.08\\
$\hat{G}_2$ & 0.19 & 0.01 & 0.00\\
\hline 
Total (\%) & 3.99 & 1.43 &  1.59 \\
\hline
\hline
\end{tabular}
\label{tab:errorbudget}
\end{table}

\begin{figure}
\includegraphics[width=0.9\hsize]{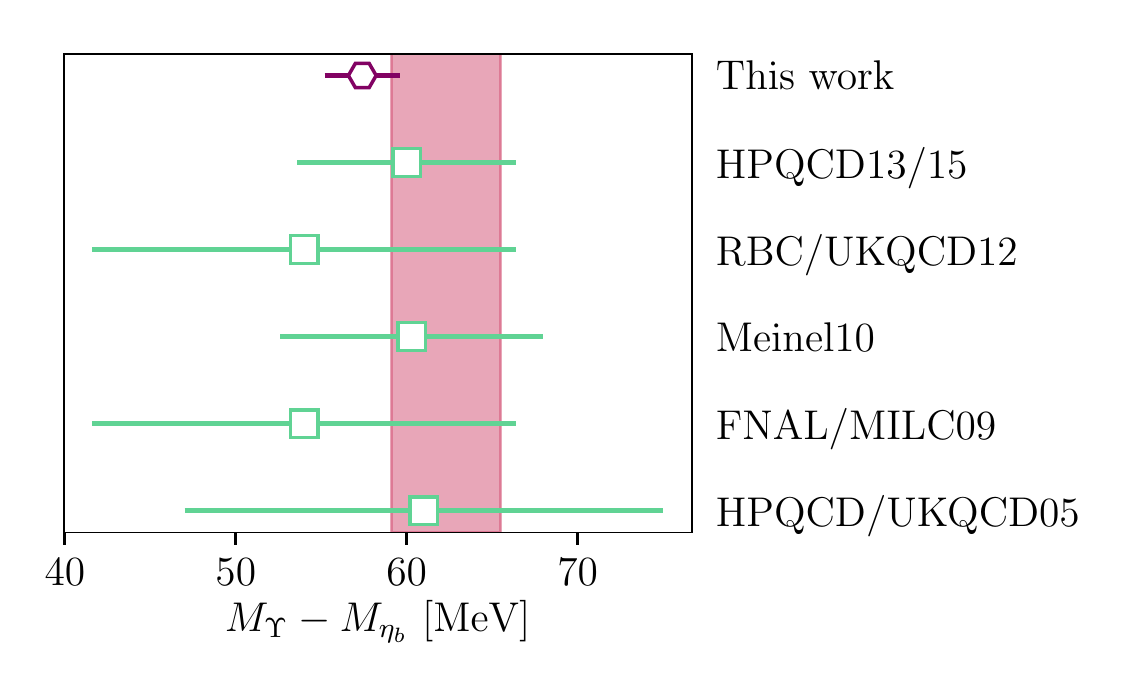}
\caption{Comparison of lattice QCD determinations of the bottomonium 
hyperfine splitting. Our result from Eq.~\eqref{eq:hypgive} is given by 
the top purple hexagon. Previous results (green squares) come from: 
HPQCD/UKQCD using $\mathcal{O}(v^4)$ NRQCD $b$ quarks and 
2+1 flavours of asqtad sea quarks~\cite{gray}; 
the Fermilab Lattice/MILC collaborations using the Fermilab formalism 
for the $b$ quark and 2+1 flavours of asqtad 
sea quarks~\cite{Burch:2009az}; 
S.Meinel using NRQCD $b$ quarks with $\mathcal{O}(v^6)$ spin-dependent 
terms and 2+1 flavours of 
domain-wall sea quarks~\cite{Meinel:2010pv}; 
the RBC/UKQCD collaboration using the 
RHQ formalism for the $b$ quark and 2+1 flavours of domain-wall 
sea quarks~\cite{Aoki:2012xaa} 
and HPQCD using 
radiatively-improved NRQCD $b$ quarks with $\mathcal{O}(v^6)$ spin-dependent 
terms and 2+1+1 flavours of HISQ 
sea quarks~\cite{Dowdall:2013jqa}. All of these results come from 
calculation of connected correlation functions and do not include an 
uncertainty from missing quark-line disconnected diagrams, 
except for~\cite{Dowdall:2013jqa}. 
\cite{Dowdall:2013jqa} includes the effect of these
disconnected diagrams through 
the inclusion of 4-quark operators with 
coefficients, calculated in perturbation theory 
through $\mathcal{O}(\alpha_s)$. See the text for discussion of 
the impact on the hyperfine splitting through $\eta_b$ 
annihilation to gluons. 
The red band is the PDG experimental average~\cite{Zyla:2020zbs}. 
The result for the hyperfine splitting calculated here 
shows a clear improvement on previous 
lattice QCD results, as well as being the first to include QED effects. 
This improvement is in large part due to the elimination of 
systematic uncertainties from the use of nonrelativistic actions 
which arise in previous calculations.}
\label{fig:hf-comp-latt}
\end{figure}

\begin{figure}
\includegraphics[width=0.9\hsize]{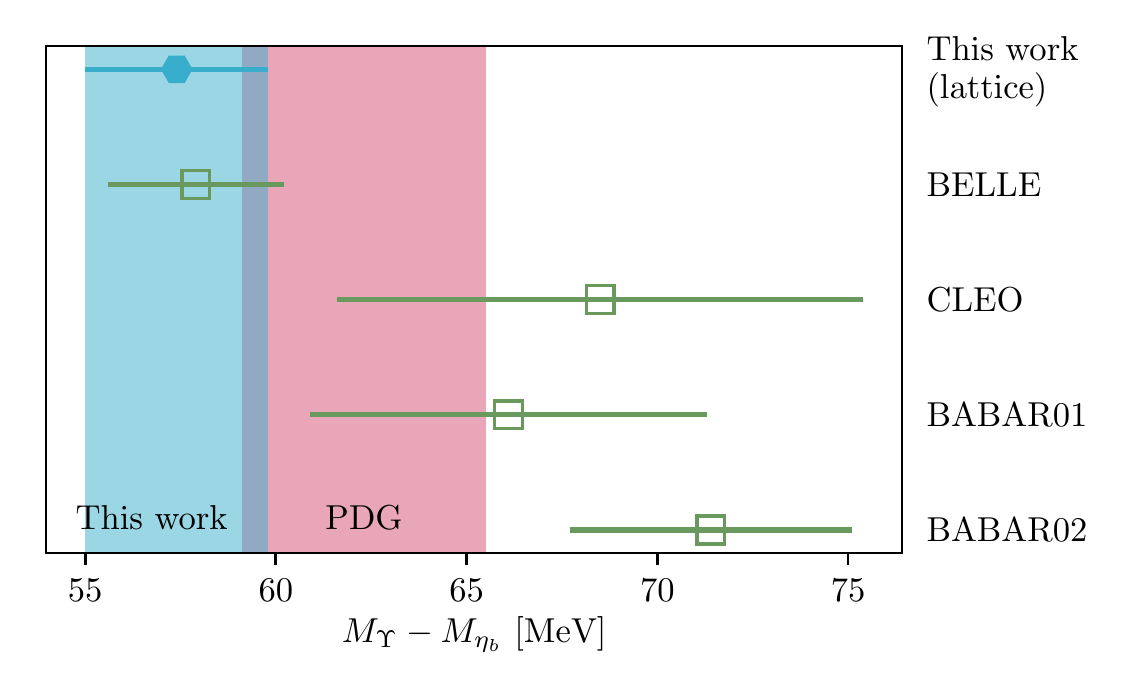}
\caption{Comparison of different experimental results for the 
bottomonium hyperfine splitting. The red band shows the PDG average 
of these experimental results~\cite{Zyla:2020zbs}. 
The filled blue hexagon is our result (Eq.~\eqref{eq:hypincdisc}) 
and is carried downwards as the blue band. Note that our result here includes 
an uncertainty from the effect of $\eta_b$ annihilation missing from 
our lattice calculation.  
There is some tension between the different experimental results 
with our value favouring the most recent result from BELLE~\cite{Mizuk:2012pb}. 
The result labelled CLEO is from \cite{Bonvicini:2009hs}, BABAR01 from 
\cite{Aubert:2009as} and BABAR02 from \cite{Aubert:2008ba}.}
\label{fig:hf-comp-exp}
\end{figure}

\subsection{Discussion: Hyperfine splitting} 
\label{sec:discusshyp}

Our bottomonium hyperfine splitting result of Eq.~\ref{eq:hypgive} is compared 
to earlier lattice QCD results in Fig.~\ref{fig:hf-comp-latt}, going back 
to the first lattice QCD calculation to include sea quarks~\cite{gray}. 
Clearly 
the use of the heavy-HISQ approach has allowed us to reduce the uncertainty 
significantly (by a factor of 3)
relative to these earlier results. The earlier results all use 
non-relativistic actions, or actions with non-relativistic input such as the 
Fermilab formalism~\cite{fermilab}, for the $b$ quarks. 
This leads to uncertainties from the normalisation of relativistic corrections 
to the Hamiltonian, such as the $\sigma\cdot B$ term that is responsible for 
the hyperfine splitting. We avoid this uncertainty with the HISQ action 
at the cost of having to calculate at multiple heavy quark masses rather than 
directly at the $b$ quark mass.  

As discussed in Sec.~\ref{sec:latt}, we have only computed 
connected correlators. This is also true for the earlier results except 
for that in~\cite{Dowdall:2013jqa}. This means that we are neglecting the contribution 
to the $\eta_b$ mass from its annihilation to gluons. This contribution 
can be related to the $\eta_b$ hadronic width using NRQCD perturbation 
theory~\cite{hisqdef}:
\begin{equation}
\Delta M_{\eta_b} = \frac{\Gamma_{\eta_b}}{2} \left( \frac{2(\mathrm{ln}2 - 1)}{\pi} + \mathcal{O}(\alpha_s,v^2/c^2) \right) .
\end{equation}
Using the total width of the $\eta_b$ of 10(5) MeV~\cite{Zyla:2020zbs} 
gives a shift to the $\eta_b$ mass from the leading order term of 
-1.0(5) MeV. 
This would result in an upward shift in the hyperfine splitting of 
approximately 1 MeV, which amounts to 0.5$\sigma$ for our 
result (Eq.~\eqref{eq:hypgive}). 

We recently showed, for the first time, 
that this leading-order analysis fails in the case of the charmonium 
hyperfine splitting~\cite{Hatton:2020qhk} where, with the improved accuracy 
we were able to achieve, it becomes clear that the lattice QCD+QED 
result is significantly higher than the experimental average. 
Assuming that this difference is the result of the effect of 
$\eta_c$ annihilation missing from the lattice calculation, 
it seems that the leading-order perturbative analysis is misleading 
in this case. Presumably missing higher-order terms in the 
perturbative analysis or nonperturbative 
effects from mixing between the $\eta_c$ and other flavour-singlet 
pseudoscalar mesons~\cite{levkova}, or both combined, have a larger effect 
than the leading-order term and opposite sign. In the bottomonium case the $\eta_b$ is 
considerably further from these lighter states and so we may expect 
a much smaller effect from this. We also expect perturbation theory to be more 
reliable at the higher energy associated with bottomonium states.

We therefore allow an additional 1 MeV uncertainty for the impact of 
$\eta_b$ annihilation on the hyperfine splitting and give a final 
result of 
\begin{equation}
M_{\Upsilon} - M_{\eta_b} = 57.5(2.3)(1.0)\ \mathrm{MeV} .
\label{eq:hypincdisc}
\end{equation}
The first uncertainty is from the lattice calculation and the second 
from missing quark-line disconnected contributions. 

The experimental average value for the bottomonium hyperfine 
splitting (62.3 $\pm$ 3.2 MeV)~\cite{Zyla:2020zbs} 
is shown by a red band on Fig.~\ref{fig:hf-comp-latt}. 
A more detailed comparison with experimental results is given in 
Fig.~\ref{fig:hf-comp-exp}. This makes clear the spread in the 
experimental results, handled in~\cite{Zyla:2020zbs} by increasing the uncertainty in the 
average by a factor of 1.8. In particular it shows that the most recent and most 
precise result from BELLE~\cite{Mizuk:2012pb} is noticeably lower than the others. 
This BELLE result is in agreement with our determination to within 
1$\sigma$. 

Our result is also in agreement with the PDG average to within 
1.5$\sigma$. We see no disagreement with the experimental result that would 
signal a larger contribution from $\eta_b$ annihilation than the 1 MeV 
that we have allowed above. Indeed a shift upwards of our hyperfine splitting 
result by 1 MeV, as suggested by leading-order perturbation theory, would 
improve the agreement between lattice QCD and experiment, 
although the shift would not be significant.
In contrast, a shift downwards of the bottomonium 
hyperfine splitting by several MeV, as we found 
for the charmonium hyperfine splitting, would cause tension with the experimental 
results. 

Finally we note that the high precision we are able to achieve for 
the bottomonium hyperfine splitting is the result of concentrating on 
the ground-state mesons with a highly-improved relativistic action. 
For a more complete picture of the bottomonium spectrum, obtained on an anisotropic lattice 
with the Fermilab heavy quark action and focussing on 
highly excited states see~\cite{Ryan:2020iog}. 

\section{$\Upsilon$ and $\eta_b$ decay constants}
\label{sec:decconsts}

We define
the vector heavyonium meson ($\phi_h$) decay constant from the 
annihilation matrix element as
\begin{equation}
\label{eq:fphidef}
\langle 0 | \overline{\psi}\gamma^i \psi | \phi_h \rangle = f_{\phi_h}M_{\phi_h} \epsilon^i \, .
\end{equation}
This means that we can determine the decay constant from our fits to the vector 
meson correlation functions using:
\begin{equation}
\frac{af_{\phi_h}}{Z_V} =  \sqrt{\frac{2A_0^V}{E_0^V}},
\label{eq:atof}
\end{equation}
where $A_0^V$ is the ground-state amplitude from a correlator fit of 
the form given in Eq.~\eqref{eq:v-corr-fit}. $Z_V$ is the renormalisation constant
required to match the local vector current in lattice QCD to
that of continuum QCD at each value of the lattice spacing.
We use $Z_V$ values calculated in a nonperturbative implementation
of the RI-SMOM scheme~\cite{Aoki:2007xm,Sturm:2009kb,Hatton:2019gha}.
The pure QCD results for $Z_V$ for the HISQ action are given in~\cite{Hatton:2019gha, Hatton:2020qhk}; we use values at scale $\mu =$ 2 GeV. 
Note that no additional matching factor is required to reach 
$\overline{\mathrm{MS}}$ 
from the RI-SMOM scheme and, because $Z_V$ has no anomalous dimensions, any 
$\mu$ dependence is purely a discretisation effect~\cite{Hatton:2019gha}. 

The vector meson decay constant is
the amplitude for annihilation
of the valence quark/antiquark pair, into a photon, for example. 
It is
related to the experimentally measurable leptonic width by:
\begin{equation}
\Gamma(\phi_h \rightarrow e^+e^-) = \frac{4\pi}{3}\alpha_{\mathrm{QED}}^2 e_h^2 \frac{f_{\phi_h}^2}{M_{\phi_h}}
\label{eq:vdecay}
\end{equation}
where $e_h$ is the quark electric charge (1/3 for $b$).
The $\alpha_{\mathrm{QED}}$ here is evaluated at 
the mass of the heavy quark and is equal to 1/132.15~\cite{Pivovarov:2000cr} 
at the $b$.

We also compute the decay constant of the pseudoscalar heavyonium 
meson, $f_{\eta_h}$. In terms of 
the parameters of our correlator fit, Eq.~\eqref{eq:ps-corr-fit} this is defined as:
\begin{equation}
\label{eq:fPdef}
f_{\eta_h} = 2m_h \sqrt{\frac{2A_0^P}{(E_0^P)^3}} .
\end{equation}
Because the partially conserved axial current (PCAC) relation holds 
for HISQ quarks the pseudoscalar decay constant is absolutely normalised 
and no $Z$ factor is required to match to the continuum regularisation of 
QCD. Since the pseudoscalar meson does not annihilate to a single particle, there 
is no experimental decay process that gives direct acces to the decay constant. 
Its value is nevertheless of interest for comparison to that of the corresponding vector 
meson and other pseudoscalar mesons.  

\begin{figure} 
\includegraphics[width=0.9\hsize]{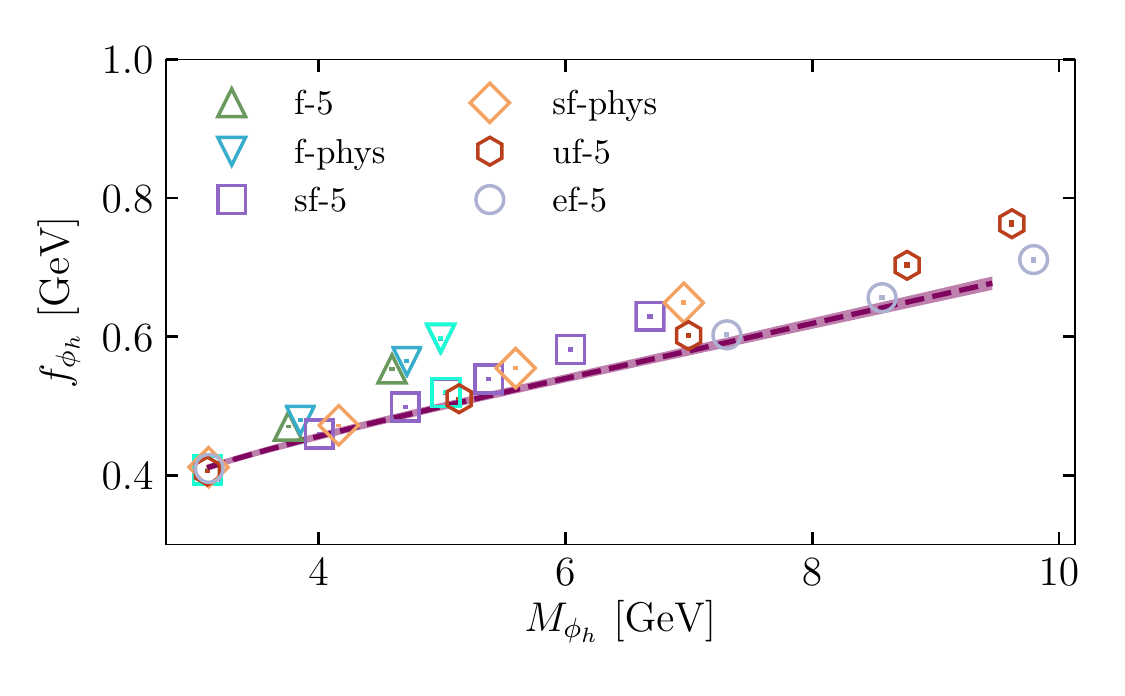}
\includegraphics[width=0.9\hsize]{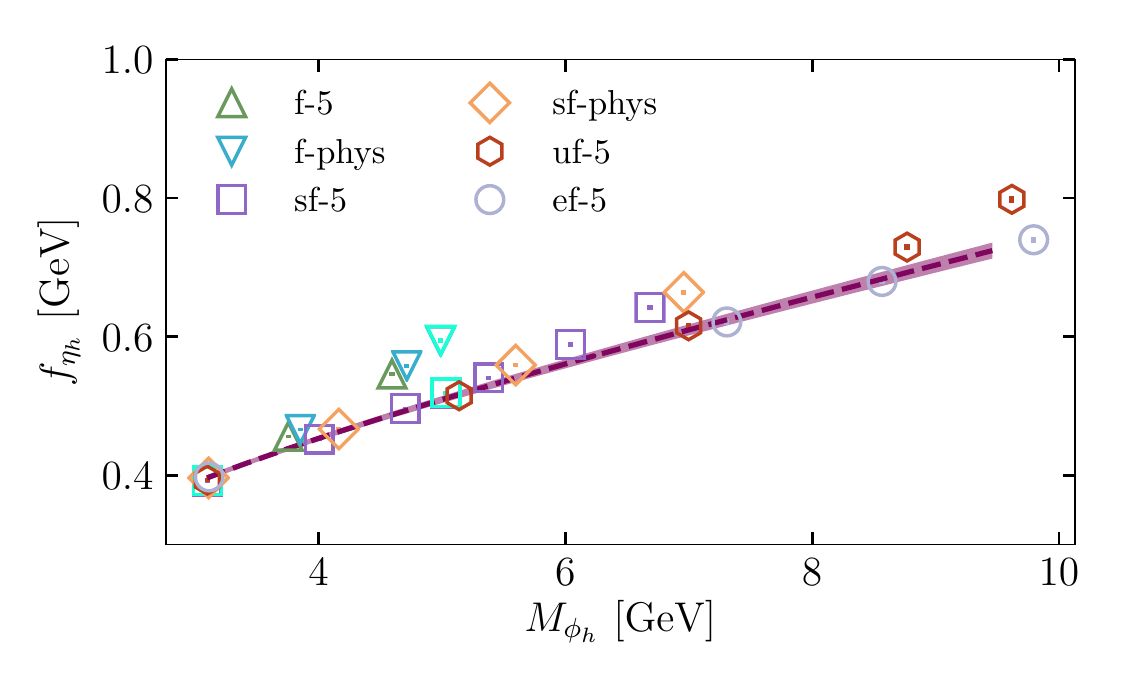}
\caption{Upper panel: The $\phi_h$ decay constant 
plotted against the $\phi_h$ mass. The symbols correspond to different gluon 
field ensembles, as given in the legend (see Table~\ref{tab:params} for a 
list). Points including quenched QED are shown in cyan, indistinguishable from 
pure QCD points underneath. The dashed line and error band show the fit described in the text evaluated at zero lattice spacing and physical sea quark masses. 
Lower panel: The $\eta_h$ decay 
constant plotted against the $\phi_h$ mass, symbols and fit line as above. }
\label{fig:decay-consts}
\end{figure}

The values of the decay constants, in lattice units, on each ensemble and for each heavy 
mass are given in the sixth and seventh columns of Table~\ref{tab:massres}.
The decay constants converted to GeV units, and renormalised in the case 
of the vector decay constant, are plotted as a function of the $\phi_h$ mass 
in Fig.~\ref{fig:decay-consts}.  The decay constants increase with increasing 
$\phi_h$ mass. Discretisation effects are clearly visible that cause the lattice 
results to peel away from the physical curve upwards. The same effect was 
seen previously for both heavy-light and heavyonium mesons~\cite{bshisq,McNeile:2012qf}. 

We also show results in Fig.~\ref{fig:decay-consts} that include the effect of quenched 
QED. Those results are given in Table~\ref{tab:qedres} as the ratio of values in QCD+QED to 
those in pure QCD. For the decay constant of the 
$\phi_h$ these ratios do not include the impact of QED on the vector current renormalisation 
factor, $Z_V$. This was calculated in~\cite{Hatton:2019gha} for the case of a quark with 
electric charge $2e/3$, again as a ratio of results in QCD+QED to those in pure QCD.  
These results are given in Table IV of~\cite{Hatton:2019gha}, with further results 
in Table X of~\cite{Hatton:2020qhk}. The ratio is within 0.05\% 
of 1, as expected for an $\mathcal{O}(\alpha_{\mathrm{QED}})$ correction to a $Z$ factor 
that is already very close to 1 for the HISQ action in pure QCD. Here we need 
results for an electric charge of $e/3$ so we determine the ratios in QCD+QED to pure QCD in 
that case by taking the values from~\cite{Hatton:2019gha, Hatton:2020qhk} 
(for $\mu= 2$ GeV)
and dividing the difference from 
1 by a factor of four.  

To fit our decay constant results as a function of lattice spacing and 
heavy quark mass we again use the fit form of Eq.~\eqref{eq:spline-fit} but 
 we use $A =$ 0.7 GeV, as appropriate for the decay constant values, and a different form 
for $F_0$ to that used for the hyperfine splitting case. The dependence 
of the decay constants on the heavy mass is approximately 
linear and so we choose $F_0 = c_F^{(0)} + c_F^{(1)} M_{\phi_h}/(3\,\mathrm{GeV})$, where 
$c_F^{(0)}$ and 
$c_F^{(1)}$ are fit parameters with prior values of $0\pm 1$.
We fit the $\phi_h$ and $\eta_h$ decay constants simultaneously, including 
the correlations between them and 
take the same $F_0$ for both since they are so close in value. The spline functions 
that map out the differences from $F_0$ in physical heavy quark mass dependence and 
the dependence on the lattice spacing and sea quark masses take independent values in the 
two cases. 
The fit has a $\chi^2/\mathrm{dof}$ value of 0.44 using an SVD cut of 
$1\times 10^{-4}$. We again evaluate our fits at zero lattice 
spacing, physical sea quark 
masses and with $M_{\phi_h}=M_{\Upsilon}$ to obtain the physical 
bottomonium results.   

We obtain, for the $\Upsilon$, 
\begin{equation}
f_{\Upsilon} = 677.2(9.7)\ \mathrm{MeV}
\label{eq:fvres}
\end{equation}
with
\begin{equation}
R_{\mathrm{QED}}[f_{\Upsilon}] = 1.00004(76) .
\end{equation}
For the $\eta_b$, 
\begin{equation}
f_{\eta_b} = 724(12)\ \mathrm{MeV} ,
\label{eq:fpsres}
\end{equation}
with 
\begin{equation}
R_{\mathrm{QED}}[f_{\eta_b}] = 1.00017(71) .
\end{equation}
Again, QED effects are not discernible within our 0.1\% uncertainties. 
At the charmonium end of our range our results agree within uncertainties 
with the values we obtained in~\cite{Hatton:2020qhk}, remembering that the calculation 
done here is for an electric charge that does not match that of the $c$ quark. 
The error budget for both decay constants is given in 
Table~\ref{tab:errorbudget}. The fit curves evaluated at zero latting 
spacing and physical sea quark masses are plotted as a function of 
heavy quark mass (given by $M_{\phi_h}$) in Fig.~\ref{fig:decay-consts}.  
The fit parameters required to reproduce these physical curves 
of the decay constants as a function of $M_{\phi_h}$ 
are given in Table~\ref{tab:vecpsdecayfit} 
of Appendix~\ref{sec:appendix}. 

\begin{figure} 
\includegraphics[width=0.9\hsize]{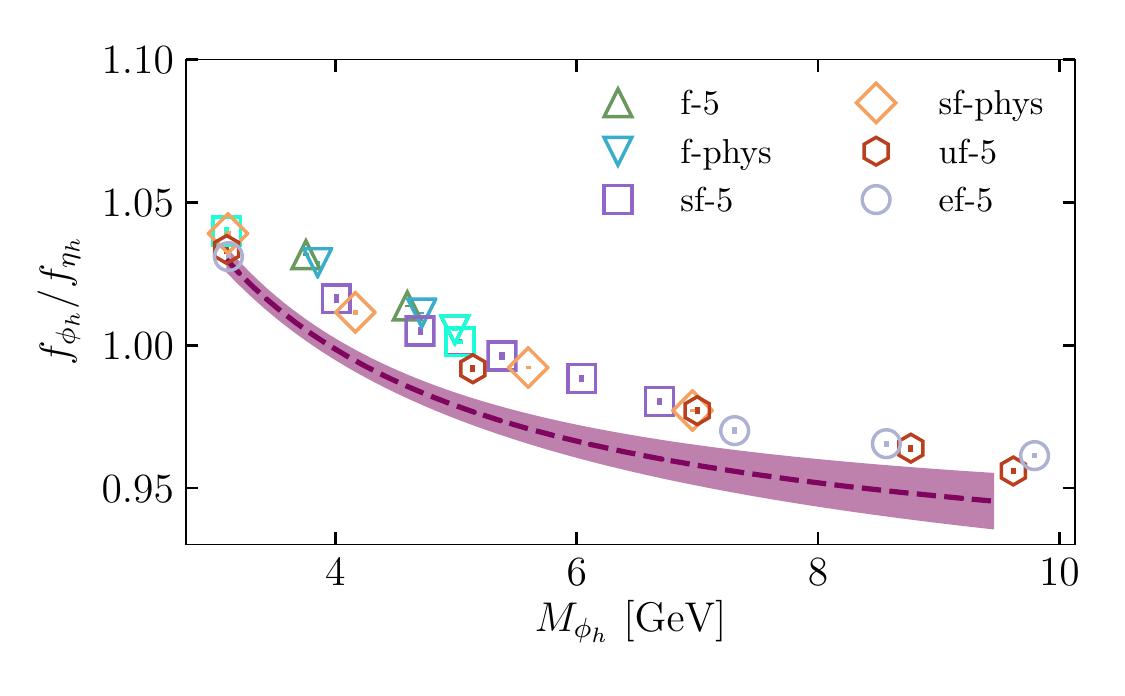}
\caption{The ratio of the vector to pseudoscalar heavyonium decay 
constants as a function of vector heavyonium mass. 
At the charmonium point the ratio is above 1. By the 
bottomonium point the ratio has shifted to be below 1. The symbols correspond 
to results on the different gluon field configurations listed in the legend 
with cyan points corresponding to QCD+QED. The line is the fit curve evaluated 
at the physical point as a function of $M_{\phi_h}$ described in the text. }
\label{fig:fVdivfP}
\end{figure}

Given that the heavy mass dependence and discretisation 
effects in the vector and pseudoscalar decay constants are 
similar we can study the ratio of the two as a function of 
the heavy mass to high precision. Our results for the ratio are 
shown as a function of $M_{\phi_h}$ in Fig.~\ref{fig:fVdivfP}. 
A slow downward drift of the ratio is seen with increasing 
$M_{\phi_h}$ from a value slightly above 1 for $c$ quarks to a value 
slightly below 1 for $b$ quarks. 

To obtain a physical result for the ratio we again use the 
fit form of Eq.~\eqref{eq:spline-fit}, now taking $F_0$ to 
be a constant, $c_F$, since the ratio is relatively flat, so 
that the spline functions handle all of the mass dependence. 
We take the prior value of $c_F$ to be 
1(1), i.e. with a very conservative width. 
Since we expect a lot of systematic effects to cancel in this ratio 
(and Fig.~\ref{fig:fVdivfP} shows that they do) we halve 
the prior widths on all of the correction terms in Eq.~\ref{eq:spline-fit} 
i.e. we take prior values on the function values at the knots of 0.0(5). 
The fit has a $\chi^2/\mathrm{dof}$ of 0.22 and no SVD cut is required.
Evaluating the fit function at the physical point gives
\begin{equation}
\frac{f_{\Upsilon}}{f_{\eta_b}} = 0.9454(99) ,
\label{eq:frat}
\end{equation}
and
\begin{equation}
R_{\mathrm{QED}}\left[\frac{f_{\Upsilon}}{f_{\eta_b}} \right] = 0.99994(38) .
\end{equation}
The total uncertainty in the ratio for the $b$ is 1\%, with a value 
clearly below 1. The fit curve evaluated at zero lattice spacing and physical sea quark masses is plotted as a function of $M_{\phi_h}$ in 
Fig.~\ref{fig:fVdivfP}.  
The fit parameters required to reproduce this physical curve 
are given in Table~\ref{tab:ratdecayfit}
of Appendix~\ref{sec:appendix}. 

\subsection{Discussion : Decay constants}

\begin{figure} 
\includegraphics[width=0.9\hsize]{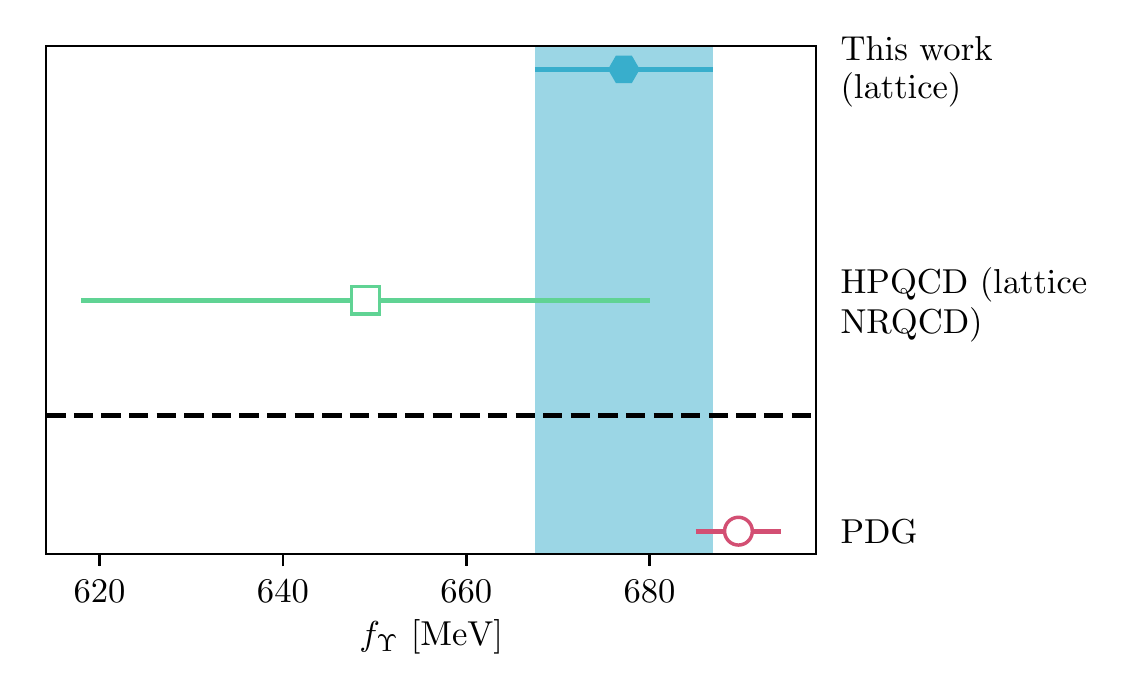}
\caption{A comparison of our result (filled blue hexagon) 
for the decay constant of 
the $\Upsilon$ with HPQCD's earlier lattice QCD result
using NRQCD $b$ quarks~\cite{Colquhoun:2014ica} (open green square). 
We also include the value inferred from the experimental 
leptonic decay width in Eq.~\eqref{eq:fupsexp} (pink open circle).  }
\label{fig:fups_comp}
\end{figure}

Figure~\ref{fig:fups_comp} compares our result for the $\Upsilon$ decay 
constant, $f_{\Upsilon}$, to that of 
an earlier lattice QCD calculation on a subset of the same gluon field 
configurations used here but using an improved NRQCD action for 
the $b$ quarks~\cite{Colquhoun:2014ica}. 
Clearly, we achieve a considerably improved 
uncertainty over that of~\cite{Colquhoun:2014ica}. A large amount 
of the NRQCD uncertainty arises from the normalisation 
and $\mathcal{O}(v^2/c^2)$ 
improvement of the NRQCD vector current, where $v$ is the nonrelativistic 
quark velocity. 
Here, since we use the HISQ action which is 
relativistic and we have performed the vector current renormalisation to 
very high precision previously~\cite{Hatton:2019gha}, these sources of uncertainty are effectively 
eliminated. 

Figure~\ref{fig:fups_comp} also compares our result for $f_{\Upsilon}$ to that obtained 
from the experimental average for the $\Upsilon$ leptonic width using 
Eq.~\eqref{eq:vdecay}. Using $\Gamma(\Upsilon \rightarrow e^+e^-)=$ 1.340(18) keV~\cite{Zyla:2020zbs} gives  
\begin{equation}
\label{eq:fupsexp}
f_{\Upsilon}^{\mathrm{expt}}=689.7(4.6)(0.8)\,\,\mathrm{MeV}
\end{equation}
where the first uncertainty comes from the experimental uncertainty 
in $\Gamma$ and the second allows for an 
$\mathcal{O}(\alpha_{\mathrm{QED}}/\pi)$ uncertainty from higher-order 
in QED terms in Eq.~\eqref{eq:vdecay} coming, for example, 
from final-state radiation. Note that using $\alpha_{\mathrm{QED}}$ of 
1/137 here instead of 1/132.15 would increase the experimental 
value of $f_{\Upsilon}$ by 3.7\% or 25 MeV. This is several times larger 
than either the experimental uncertainty or our lattice QCD uncertainty.  

Figure~\ref{fig:fups_comp} shows good agreement, within 1$\sigma$, 
 between our lattice QCD result and 
that from experiment (eq.~\eqref{eq:fupsexp}). 
The experimental uncertainty is about half that from 
our lattice QCD result. 

Our result for $f_{\Upsilon}$ can be converted into a determination 
of the width for $\Upsilon$ decay to light leptons in the Standard 
Model using Eq.~\eqref{eq:vdecay}. 
This gives 
\begin{equation}
\label{eq:ourgamma} 
\Gamma(\Upsilon \rightarrow e^+e^-) = 1.292(37)(3) \, \mathrm{keV} 
\end{equation}
where the first uncertainty comes from the lattice QCD result and 
the second allows for a relative $\mathcal{O}(\alpha_{\mathrm{QED}}/\pi)$ 
correction to Eq.~\eqref{eq:vdecay} from higher-order QED effects. 

Our result for $f_{\eta_b}$ can be compared to an earlier HPQCD 
lattice QCD result 
using HISQ quarks and the heavy-HISQ approach 
on gluon field configurations including the 
effect of 2+1 flavours of asqtad sea quarks~\cite{McNeile:2012qf}.  
That work obtained a value $f_{\eta_b} =$ 667(6) MeV, 
which is significantly lower (by 4$\sigma$) than our result here.  
The discrepancy is most likely to result from 
a bias in the earlier results from not having values on lattices 
with spacings as fine as we do here. 
Another possible source of the discrepancy is the fact 
that the earlier calculation 
did not include $c$ quarks in the sea. 
Having more flavours of quarks in the sea 
results in a slower running of the strong coupling constant. Hence, using 
the language of potential models, we expect the 
Coulomb-like term in the heavy quark 
potential (of the form $-4\alpha_s(r)/(3r)$) to have a larger value 
for $\alpha_s$ at the short-distance scales to which the $\eta_b$ meson 
decay constant is sensitive. This corresponds to a deeper potential 
at short distances and a correspondingly larger `wavefunction-at-the-origin', 
which is the quantity in a potential model that translates approximately into
the decay constant. This effect could explain some of the discrepancy but is 
unlikely to be large enough to explain it all. 
The calculations in~\cite{McNeile:2012qf} also used a different form to fit 
the lattice results as a function of heavy quark mass (in that case using 
as proxy $M_{\eta_h}$). This consisted of multiple powers of the inverse 
heavy quark mass multiplied by a leading function of the form $(M/M_0)^b$
where $b$ was allowed to float. We have checked that using that fit 
form here gives us results for $f_{\eta_b}$ very consistent with our 
spline fits, so the discrepancy with~\cite{McNeile:2012qf} is not related 
to the form of the fit used. 

The ratio of vector to pseudoscalar decay constants as a function 
of heavyonium mass provides a test of our understanding of these 
mesons. In the language of potential models the heavyonium vector 
and pseudoscalar mesons differ only through spin-dependent relativistic 
corrections to the central potential~\cite{Davies:1997hv}. The size of 
relativistic corrections fall as the heavy quark mass increases and 
the mean squared-velocity of the heavy quarks fall. In the infinite 
quark mass limit pseudoscalar and vector heavyonium mesons have the same 
mass and the same wavefunction-at-the-origin. The decay constants differ, 
however, by the matching factors that are needed to renormalise 
temporal axial and spatial vector currents from this nonrelativistic 
framework to full continuum QCD. 
The ratio of the vector to pseudoscalar 
heavyonium decay constants would then be expected 
to become the ratio of the vector to temporal axial vector matching 
factors in the heavy quark limit. 
The matching factors come from 
high-momentum regions of phase-space and so can be calculated 
in QCD perturbation theory. An $\mathcal{O}(\alpha_s)$ matching 
 calculation was done in~\cite{Jones:1998ub} for spin-independent 
nonrelativistic QCD and gave the result 
\begin{equation}
\label{eq:Zrat}
\frac{Z_V}{Z_A} = 1-\frac{g^2}{6\pi^2} = 1 - \frac{2\alpha_s}{3\pi} .
\end{equation} 
From this we conclude that we would expect the ratio of vector to pseudoscalar 
decay constants to be below 1 for large heavy quark mass. 
Eq.~\eqref{eq:Zrat} expects the difference from 1 to be $\mathcal{O}(5\%)$, 
taking $\alpha_s \approx 0.25$,  
but this formula will have corrections from higher orders in $\alpha_s$. 
A value for the ratio of 5\% below 1 is very consistent with our 
results in Fig.~\ref{fig:fVdivfP}, however. 

Very similar behaviour is seen for the ratio of vector to pseudoscalar 
decay constants for heavy-light mesons from lattice QCD calculations. 
The decay constant of the $D_s^*$ meson is found to be 
several percent larger than that 
of the $D_s$~\cite{Donald:2013sra, Lubicz:2017asp, Chen:2020qma} whereas 
that of the $B_s^*$ is a few percent below that of the 
$B_s$~\cite{Colquhoun:2015oha}~\cite{Lubicz:2017asp}. This behaviour can be 
understood on the same basis as the arguments for heavyonium above. In 
the heavy-light case an $\alpha_s^3$ calculation of the matching factors 
is available in the infinite heavy quark mass 
limit~\cite{Bekavac:2009zc}. The corrections to 
the $\mathcal{O}(\alpha_s)$ formula for the ratio (which is the same 
as for heavyonium in Eq.~\eqref{eq:Zrat}) are sizeable 
but have the same (negative) sign and so do not change the qualitative 
behaviour of the difference of the ratio from 1.  

\begin{figure} 
\includegraphics[width=0.9\hsize]{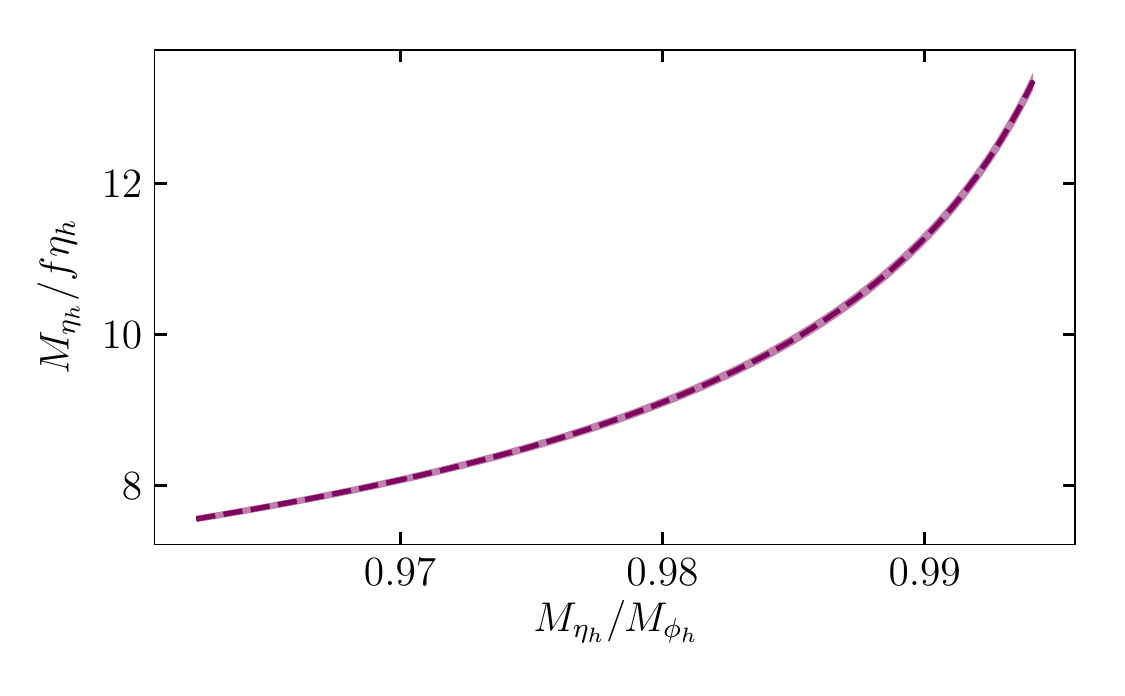}
\includegraphics[width=0.9\hsize]{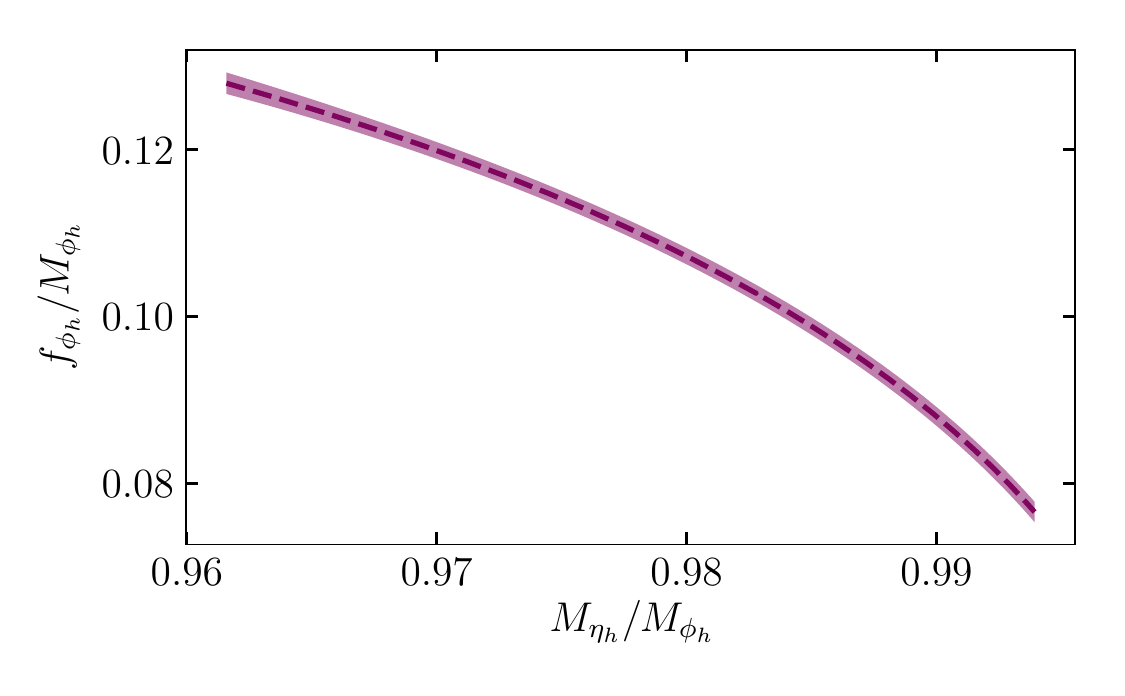}
\caption{Upper plot: The physical ratio (evaluated at zero lattice spacing and with 
sea quark mass mistunings set to zero) of the mass to decay constant for the 
pseudoscalar heavyonium meson, $\eta_h$ as a function of the ratio of pseudoscalar 
to vector heavyonium masses. 
Lower plot: The physical ratio of decay constant to mass for the vector heavyonium 
meson, $\phi_h$, (the quantity denoted $f_V$ in~\cite{DeGrand:2019vbx}) 
plotted against the same ratio of masses. 
}
\label{fig:Mdivf}
\end{figure}

Having performed the fits of the previous subsections we now 
have physical values for the decay constants not only at 
the $b$ quark mass but at the full range of masses between 
the $c$ and $b$ quark masses. The physical curves as a function of 
meson mass in Figs.~\ref{fig:decay-consts} and~\ref{fig:fVdivfP} 
could be used to tune phenomenological QCD potential models, that 
often differ markedly on features of heavyonium physics such as 
details of the wavefunction even when reproducing the spectrum 
(see, for example~\cite{kwongrosner,eichtenquigg,eichtenrosner}).  

They may also be useful beyond QCD. In~\cite{DeGrand:2019vbx} 
lattice QCD results across a range of masses were collected 
with the intention of providing useful information for 
phenomenologists studying strongly coupled beyond the Standard 
Model (BSM) theories. These theories are often QCD-like but typically 
with heavier (relative to the confinement scale) 
fundamental fermions than the light quarks of QCD. 
Ref.~\cite{DeGrand:2019vbx} makes the point that information from 
lattice QCD calculations about how (for example) 
meson masses and decay constants depend on quark masses 
can be useful to constrain such BSM theories. This then requires lattice 
QCD results for quark masses not at their physical values, as we have here. 
The lattice QCD results need to be presented in an 
appropriate way with dimensionless combinations of 
decay constants and masses on both axes. A convenient $x$-axis is the 
ratio of pseudoscalar to vector meson 
mass. In~\cite{DeGrand:2019vbx} the square of this quantity was used 
since the lattice QCD results were concentrated at light quark masses.    
Here, since we have heavy quarks and the ratio of pseudoscalar to vector 
meson masses is close to 1, we simply use the ratio. 

Dimensionless ratios are readily 
obtained for our raw lattice results using the values 
in Table~\ref{tab:massres}. Correlations can be ignored because 
statistical uncertainties are so small.  
In the following we construct appropriate ratios from our fit 
functions in the limit of zero lattice spacing and physical sea 
quark masses and do not include the raw lattice results in 
the figures, for clarity. 
 
One useful quantity~\cite{Hochberg:2014kqa} is the ratio of the pseudoscalar 
meson mass and decay constant for a meson made of quarks of 
degenerate mass (i.e. the `pions' of the BSM model). 
Using the physical heavy mass dependence 
of $f_{\eta_h}$ extracted from our fit we display the ratio 
of $M_{\eta_h}$ and $f_{\eta_h}$ as a function of the ratio of pseudoscalar 
to vector meson masses 
in Fig.~\ref{fig:Mdivf}. 
Our results show values of $M_{\eta_h}/f_{\eta_h}$ around 10, and 
continuing to rise, as the ratio of pseudoscalar to vector meson 
masses heads towards 1. Note that our definition of the 
pseudoscalar decay constant in Eq.~\eqref{eq:fPdef} corresponds to 
the normalisation $f_{\pi} \approx$ 130 MeV.   

As discussed in~\cite{DeGrand:2019vbx} composite models of a dark 
sector in which a `dark $\rho$' meson couples to ordinary matter through 
a dark photon (e.g.~\cite{Harigaya:2016rwr}) 
need information on the vector meson decay constant for 
an appropriate range of fermion masses. The ratio of vector meson decay 
constant to vector meson mass is denoted $f_V$ in~\cite{DeGrand:2019vbx}. 
In our convention for the vector meson decay constant  
(Eq.~\eqref{eq:fphidef}) it is $f_{\phi_h}/M_{\phi_h}$. 
We plot $f_{\phi_h}/M_{\phi_h}$ against the pseudoscalar to 
vector meson mass ratio in Fig.~\ref{fig:Mdivf}. We see that 
this ratio becomes small as the ratio of pseudoscalar to vector 
meson masses heads towards 1. 
It also has relatively strong dependence on the mass ratio, so 
using an approximately constant value 
(based, for example, on naive dimensional analysis) 
would not agree well with our results. 

We also note that accurate lattice QCD results are available at 
the ratio of pseudoscalar to vector meson masses of 0.673 which corresponds 
to $s\overline{s}$ mesons when only connected correlation 
functions are calculated. This means that the pseudoscalar meson is not 
allowed to annihilate and mix with flavour-singlet mesons made from 
lighter quarks, and likewise the vector decay to two pseudoscalar mesons 
incorporating lighter quarks is not included. This is then the scenario 
that would match that required in a composite BSM scenario. For 
this case HPQCD calculates
$M_{\eta_s}/f_{\eta_s}$ = 3.801(16)~\cite{fkpi} and 
$f_{\phi_s}/M_{\phi_s}$ = 0.233(3)~\cite{Chakraborty:2017hry}. These results 
must connect smoothly to the ones shown in Fig.~\ref{fig:Mdivf} as the 
quark mass is increased.

\begin{table*}
\caption{ Results in lattice units for time moments of the 
vector heavyonium correlator as defined in Eq.~\eqref{eq:timemomv1}. 
We give raw results here in which the vector current has not been renormalised and 
we also take the $(n-2)$th root to reduce all the moments to the same dimensions. 
The numbers in the table are then $(G_n^V/Z_V^2)^{1/(n-2)}$ for $n$=4 to 10. 
 }
\begin{tabular}{llllllllll}
\hline
\hline
Set & $am_h$ & $n=4$ & $n=6$  & $n=8$ & $n=10$ \\
\hline
1 & 0.6 & 0.562768(11) & 1.263877(18) & 1.849015(25) & 2.386587(33) \\
  & 0.8 & 0.4342507(59) & 1.0316257(96) & 1.525578(12) & 1.967383(15) \\
\hline
2 & 0.6 & 0.5628101(56) & 1.2640282(83) & 1.849341(10) & 2.387158(12) \\
 & 0.8 & 0.4342571(32) & 1.0316599(48) & 1.5256620(57) & 1.9675445(65) \\
 & 0.866 & 0.395358(44) & 0.966054(54) & 1.439945(54) & 1.861205(52) \\
\hline
3 & 0.274 & 1.070712(58) & 2.27651(10) & 3.35545(14) & 4.37419(17) \\
 & 0.4 & 0.797940(92) & 1.72744(16) & 2.54252(21) & 3.31432(25) \\
 & 0.5 & 0.665034(57) & 1.466318(96) & 2.15508(13) & 2.80305(16) \\
 & 0.548 & 0.60993(22) & 1.36481(25) & 2.00869(24) & 2.61113(24) \\
 & 0.6 & 0.569093(37) & 1.282926(63) & 1.886184(84) & 2.44653(10) \\
 & 0.7 & 0.495371(26) & 1.145651(43) & 1.689190(58) & 2.186270(72) \\
 & 0.8 & 0.436182(18) & 1.037685(31) & 1.538234(41) & 1.989299(51) \\
\hline
4 & 0.260 & 1.114660(44) & 2.366266(78) & 3.48827(11) & 4.54699(14) \\
 & 0.4 & 0.798236(18) & 1.728246(32) & 2.544009(44) & 3.316608(55) \\
 & 0.6 & 0.5691755(75) & 1.283151(13) & 1.886612(17) & 2.447216(22) \\
 & 0.8 & 0.4362111(38) & 1.0377647(63) & 1.5383891(83) & 1.989559(11) \\
\hline 
5 & 0.194 & 1.431378(91)  & 3.03675(16) & 4.49434(22)  & 5.86769(29) \\
 & 0.4 & 0.808461(20)  & 1.757499(36) & 2.597493(51)  & 3.396810(65) \\
 & 0.6 & 0.5722526(77) & 1.292481(13) & 1.904940(19) & 2.476827(24) \\
 & 0.8 & 0.4371741(39) & 1.0407975(65) & 1.5447616(86) & 2.000643(11) \\
 & 0.9 & 0.3876830(29) & 0.9510722(49) & 1.4215906(63) & 1.8418824(80) \\
\hline
6 & 0.138 & 1.91475(23) & 4.06357(42) & 6.02429(55) & 7.86806(66) \\
 & 0.45 & 0.739093(19) & 1.623965(33) & 2.403967(45) & 3.147423(57) \\
 & 0.55 & 0.621096(13) & 1.389882(21) & 2.052994(29) & 2.679049(36) \\
 & 0.65 & 0.5342576(87) & 1.222777(15) & 1.806579(20) & 2.349544(24) \\
\hline
\hline
\end{tabular}
\label{tab:moments}
\end{table*}

\begin{table*}
\caption{ Quenched QED corrections, for quark electric charge $e/3$, 
to the time-moments given for a subset of the results 
in Table~\ref{tab:moments} for the $(n-2)$th root of the unrenormalised $G_n^V/Z_V^2$. 
The results are given as the ratio, $R^0$, of the value in 
QCD+QED to that in pure QCD at fixed valence quark mass in lattice units.  
 }
\begin{tabular}{llllllllll}
\hline
\hline
Set & $am_h$ & $n=4$ & $n=6$  & $n=8$ & $n=10$ \\
\hline
2 & 0.866 & 0.999669(32) & 0.999731(16) & 0.999697(11) & 0.999641(8) \\
\hline
3 & 0.274 & 0.999774(30) & 0.999692(25) & 0.999646(24) & 0.999622(24) \\
 & 0.548 & 0.999475(43) & 0.999353(22) & 0.999194(14) & 0.999060(11) \\
\hline
\hline
\end{tabular}
\label{tab:qedmoments}
\end{table*}

\section{Vector current-current correlator time-moments and $a_{\mu}^b$}
\label{sec:time-moms}

The ground-state vector heavyonium decay constant is determined 
by the amplitude of the 
state that dominates the correlator at large times and this can be
connected to experiment via the leptonic width, as we have seen. 
We can also calculate the time moments of the correlator. 
These depend on the behaviour of the correlator at shorter 
time distances and can also be connected to experimental results~\cite{firstcurrcurr,psipaper}. 
The moments of the vector heavyonium correlator are defined by:
\begin{equation}
G_n^{V} = Z_V^2\sum_{\tilde{t}} \tilde{t}^n C_{\phi_h}(\tilde{t})
\label{eq:timemomv1}
\end{equation}
where $\tilde{t}$ is lattice time 
symmetrised around the centre of the lattice, $C_{\phi_h}$ is 
the vector two-point correlation function and $Z_V$ is the 
renormalisation factor for the heavyonium vector current operator used. 

\begin{figure}[h!]
\includegraphics[width=0.8\hsize]{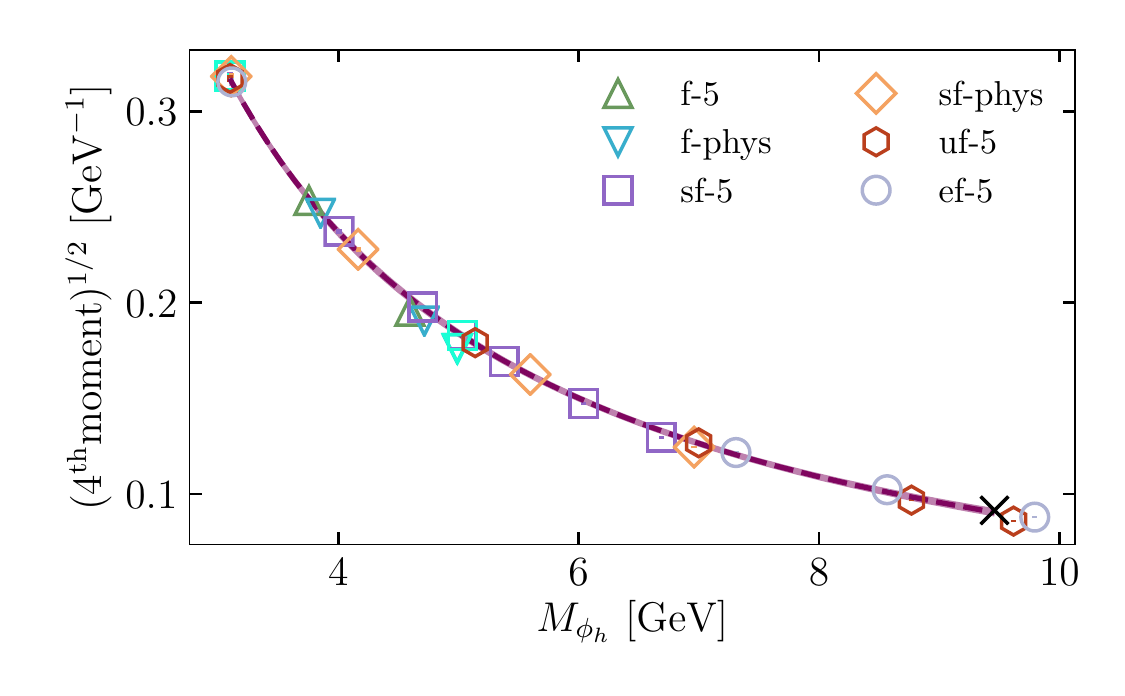}
\includegraphics[width=0.8\hsize]{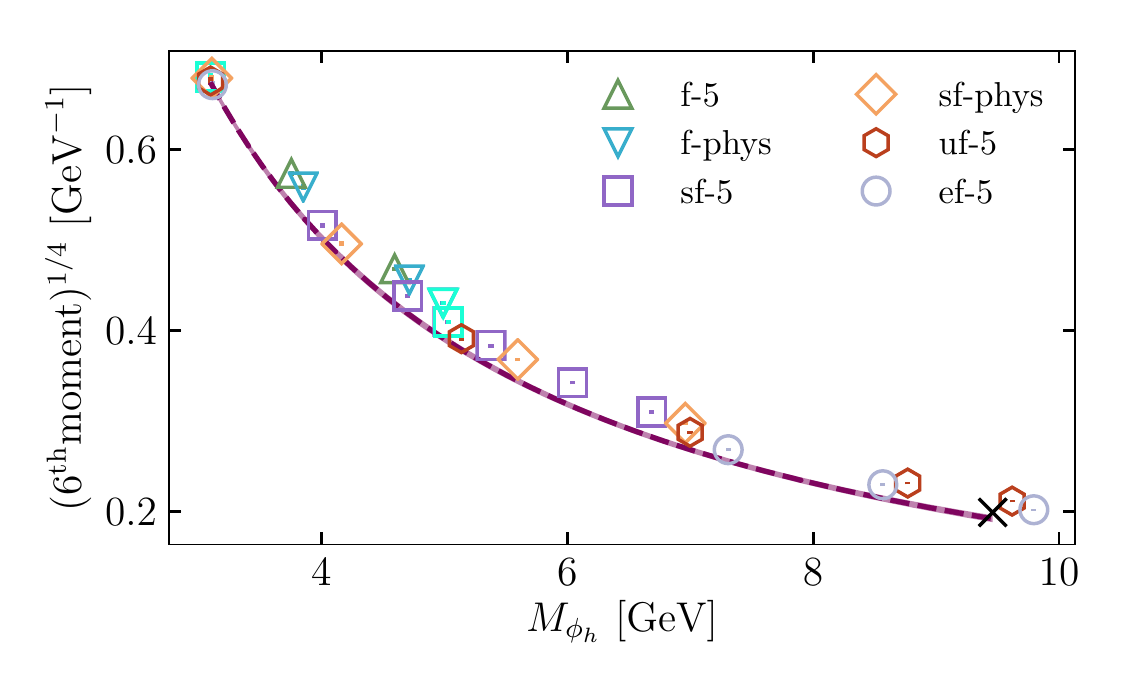}
\includegraphics[width=0.8\hsize]{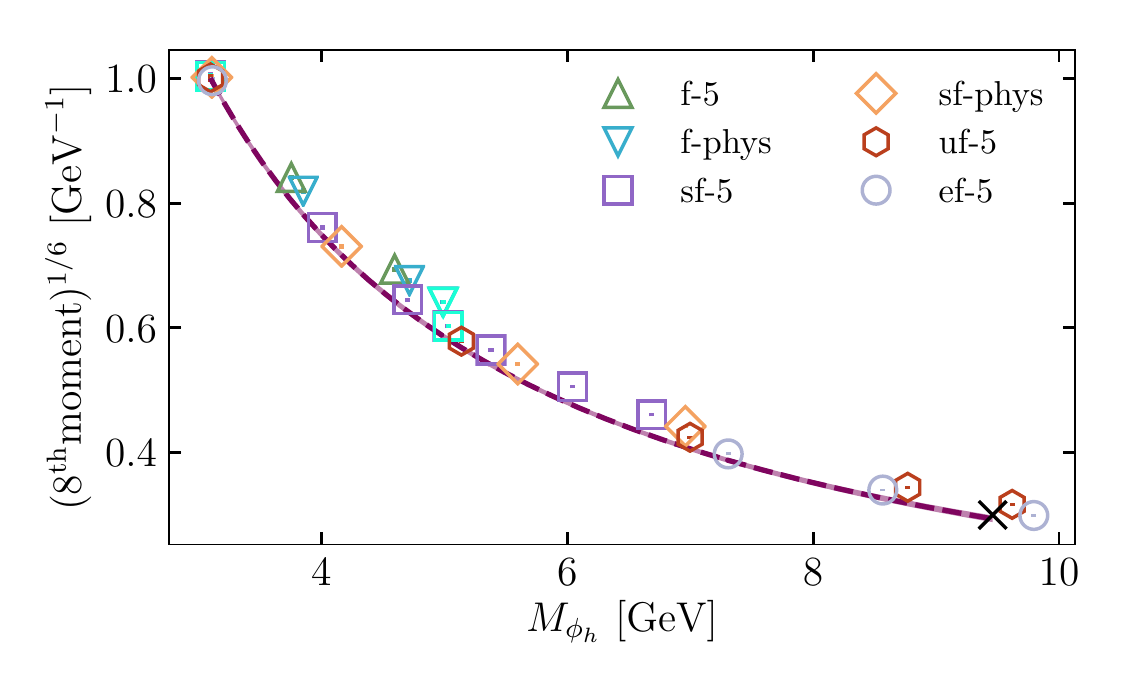}
\includegraphics[width=0.8\hsize]{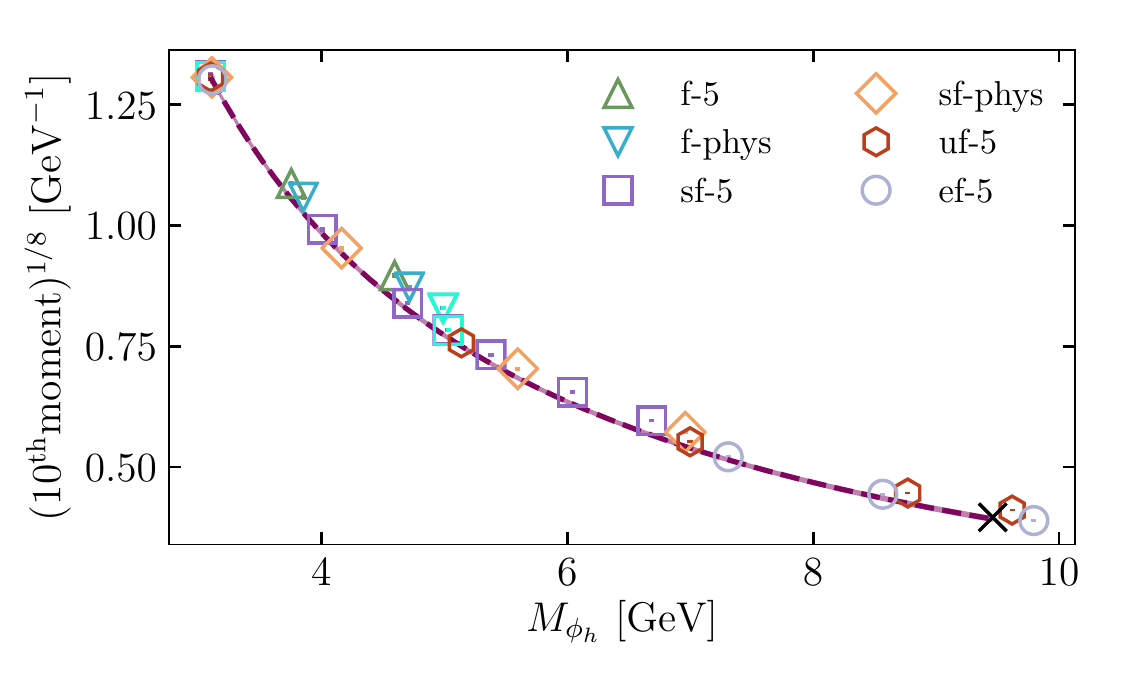}
\caption{Results for the 4th, 6th, 8th and 10th 
time moments of the heavyonium vector correlator
plotted as a function of $M_{\phi_h}$. 
The symbols correspond to different gluon field ensembles, as 
given in the legend (see Table~\ref{tab:params} for a list).
The errors on the points
are dominated by uncertainties from the 
determination of $Z_V$ that are
correlated between the points. 
Points including quenched QED are shown in cyan, indistinguishable from pure 
QCD points underneath. 
The dashed line with 
purple error band displays our continuum/chiral fit, as discussed in the text. 
Values determined from experimental results for $R_{e^+e^-}$ (eq.~(\ref{eq:rnexp}))
are plotted as the black crosses
at $M_{\phi_h}=M_{\Upsilon}$~\cite{kuhnmc07}.
}
\label{fig:rn}
\end{figure}

Results for $(G^V_n/Z_V^2)^{1/(n-2)}$ in lattice units 
on each of our ensembles are given in Table~\ref{tab:moments} 
for $n=4$ to 10. The power $1/(n-2)$ is taken to 
reduce all the moments to the same dimension. 
We take the $Z_V$ factor for the vector current to be the same one used 
for the leptonic width above~\cite{Hatton:2019gha}.
Figure~\ref{fig:rn} then shows the physical results for
these moments as a function of $M_{\phi_h}$.

Results that include quenched QED corrections for a subset 
of ensembles are given in Table~\ref{tab:qedmoments}.
These are given as the ratio of the result in QCD+QED to that in 
QCD at fixed valence quark mass in lattice units. 
The values of $R^0$ are very slightly below 1, as for charmonium~\cite{Hatton:2020qhk}.
The difference from 1 is even smaller here because of the smaller quark 
electric charge. 
Note that 
the vector current is not renormalised in these raw results and 
QED effects in $Z_V$ must also be taken into account, as for the 
decay constant~\cite{Hatton:2019gha}. 
These results are also plotted in Fig.~\ref{fig:rn} 
as the cyan points. The impact of QED is not visible.  

To fit the time-moment results as a function of lattice spacing and heavy 
quark mass we again use the fit of Eq.~\eqref{eq:spline-fit}, 
supplemented with QED effects in Eq.~\eqref{eq:spline-fit-qed}. 
For the time-moments we use $F_0 = (c_F^{(0)} + c_F^{(1)}(3\,\mathrm{GeV})/M_{\phi_h})$, 
and the dimensionful parameter $A$ is taken as 0.5 $\mathrm{GeV}^{-1}$ 
for every moment. The prior values on $c_F^{(0)}$ and $c_F^{(1)}$ 
are taken to be $0\pm 1$ for each moment.  
We fit all moments separately using an SVD cut 
of $5 \times 10^{-4}$ in all cases. The $\chi^2/\mathrm{dof}$ 
of the fits, in order of increasing $n$, are 
0.9, 0.19, 0.26 and 0.4.
The curves in Fig.~\ref{fig:rn} show the fit results 
evaluated at zero lattice spacing and with tuned sea quark masses. 

Table~\ref{tab:momfinal} gives our results for the time-moments evaluated at the 
$b$ quark mass in the continuum limit, with their total uncertainties. 
The corresponding error budget is given in Table~\ref{tab:momerrorbudget}. 
In the next section we compare these results to earlier lattice analyses and 
values determined from experimental data for $R(e^+e^- \rightarrow \mathrm{hadrons})$. 
We will also use the results to improve the determination of the 
$b$ quark contribution to the hadronic vacuum polarisation term in the 
Standard Model determination of the anomalous magnetic moment of the muon. 
Column 3 of Table~\ref{tab:momfinal} gives the ratio of the QCD+QED result 
to that in pure QCD for each moment. Again we are not able to distinguish 
any impact of QED on the results at the level of our uncertainties (which range 
from 0.4\% down to 0.1\%).  

\begin{table}
\caption{Results for the time moments of the bottomonium vector current-current correlator 
obtained from evaluating our fit functions in the continuum limit at the $b$ quark 
mass. These are given in the second column for moment numbers listed in the first column.  
The results extracted from experimental 
data in~\cite{kuhnmc07} are given in the third column for comparison. The 
fourth column gives the quenched QED correction to these moments, as a ratio of 
the value in QCD plus QED to that in pure QCD with a tuned $b$ quark mass (to reproduce 
the $\Upsilon$ mass from experiment) in both 
cases. All of the ratios
are consistent with 1.0.}
\begin{tabular}{llll}
\hline \hline
$n$ & $G_n^{1/(n-2)}$ & $\left(G_n^{\mathrm{exp.}} \right)^{1/(n-2)}$ & $R_{\mathrm{QED}} \left[ G_n^{1/(n-2)} \right]$ \\
& $[\mathrm{GeV}^{-1}]$ & $[\mathrm{GeV}^{-1}]$ & \\ 
\hline
4 & 0.0905(23) & 0.09151(31) & 0.9996(38) \\
6 & 0.1920(39) & 0.19910(49) & 0.9999(19) \\
8 & 0.2934(55) & 0.29964(55) & 0.9999(13) \\
10 & 0.3918(66) & 0.39548(59) & 0.9999(10) \\
\hline \hline
\end{tabular}
\label{tab:momfinal}
\end{table}

\begin{table}
\caption{Error budget for the $n$th time-moment, $G_n^{1/(n-2)}$, 
as a percentage of the final 
answer.}
\begin{tabular}{lcccc}
\hline
\hline
$n$ &  4 & 6 & 8 & 10 \\
\hline
statistics & 0.25 & 0.27 & 0.27 & 0.24 \\
SVD cut & 1.84 & 1.63 & 1.50 & 1.34 \\
$w_0$ & 0.59 & 0.62 & 0.62 & 0.58 \\
$w_0/a$ & 0.39 & 0.31 & 0.33 & 0.23 \\
$Z_V$ & 0.13 & 0.04 & 0.02 & 0.01 \\
$F_0$ & 0.01 & 0.02 & 0.02 & 0.03 \\
$G_0$ & 0.01 & 0.01 & 0.01 & 0.02 \\
$G_1$ & 0.75 & 0.35 & 0.26 & 0.28 \\
$G_2$ & 0.42 & 0.31 & 0.24 & 0.42 \\
$G_3$ & 0.43 & 0.14 & 0.14 & 0.19 \\
$G_4$ & 0.57 & 0.36 & 0.32 & 0.31 \\
$G_5$ & 0.97 & 0.71 & 0.67 & 0.42 \\
$\hat{G}_1$ & 0.34 & 0.18 & 0.12 & 0.09 \\
$\hat{G}_2$ & 0.01 & 0.00 & 0.00 & 0.00 \\
\hline 
Total (\%) & 2.54 & 2.03 &  1.87 &  1.68 \\
\hline
\hline
\end{tabular}
\label{tab:momerrorbudget}
\end{table}

\subsection{Discussion: time moments and $a_{\mu}^b$}
\label{sec:discussmoments}

\begin{figure}
\includegraphics[width=0.9\hsize]{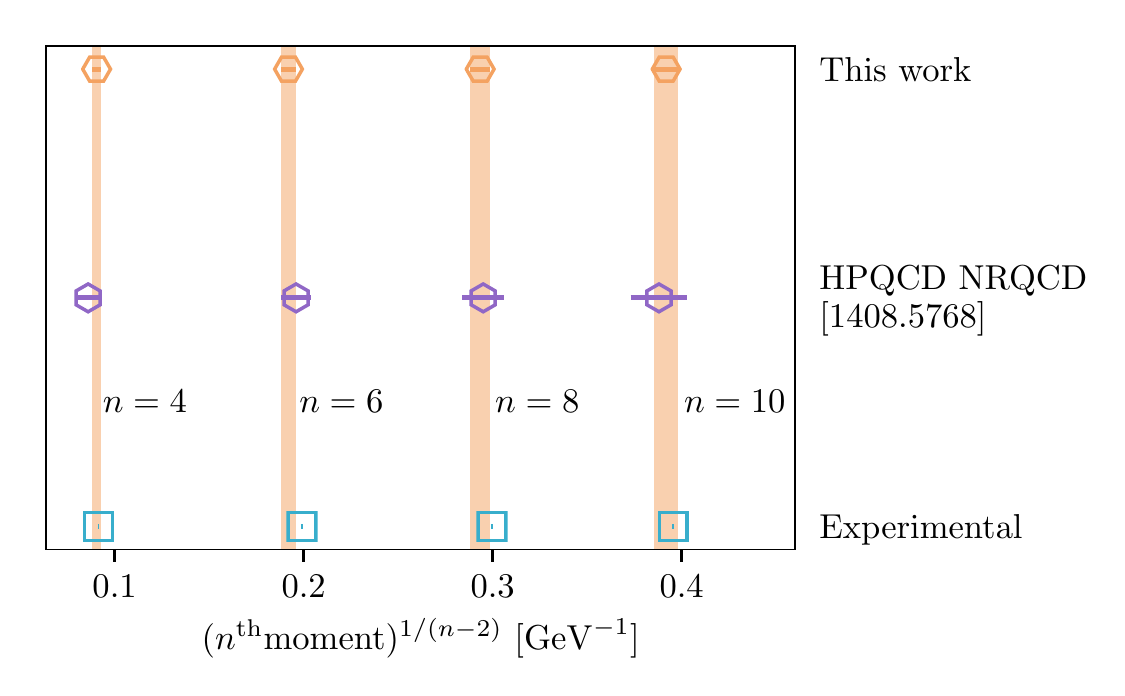}
\caption{Comparison of different determinations of the 
four lowest time moments of the bottomonium vector current-current correlator. 
The three determinations are, from the top, this work, 
previous calculation by the HPQCD collaboration using NRQCD $b$ 
quarks~\cite{Colquhoun:2014ica} and the 
values obtained from experimental data on $R(e^+e^- \rightarrow \mathrm{hadrons})$ 
in~\cite{kuhnmc07}.}
\label{fig:tmoms-comp}
\end{figure}
 
In Fig.~\ref{fig:tmoms-comp} we compare our results 
for the time-moments to those of an earlier
HPQCD calculation that used NRQCD $b$ quarks~\cite{Colquhoun:2014ica}. 
For the NRQCD results the key sources of error were from 
the vector current normalisation (using a method based on matching the 
time-moments to continuum perturbation theory) 
and from the lattice spacing dependence 
effects in the NRQCD action. 
Our uncertainties here are a 
considerable improvement (by over a factor of two) 
on the NRQCD results, because we have 
a very accurate vector current normalisation 
and have results over a large range of lattice spacing values to control 
the lattice spacing dependence. 

Figure~\ref{fig:tmoms-comp} shows that 
our results agree within 2$\sigma$ with the values extracted for 
the $q^2$-derivative moments, $\mathcal{M}_k$ ($n=2k+2$), 
of the $b$ quark vacuum 
polarization using experimental values for 
$R_{e^+e^-} = \sigma(e^+e^- \rightarrow \mathrm{hadrons})/\sigma_{pt}$~\cite{kuhnmc07}. 
The appropriate normalisation of these results for the comparison to ours, is:
\begin{equation}
G_n^{\mathrm{exp}} = \left( \frac{\mathcal{M}^{\mathrm{exp}}_k n!}{12\pi^2 e_b^2} \right)^{1/(n-2)} 
\label{eq:rnexp}
\end{equation}
Our results from lattice QCD have considerably larger
uncertainties than those of the experimental values 
but together these results 
provide a further test of QCD at the level of 2\%. 

We may also use these time-moments to extract the 
$b$ quark connected contribution to the leading order 
hadronic vacuum polarisation contribution to the 
anomalous magnetic moment of the muon. This was done 
in \cite{Colquhoun:2014ica} and, given that we have 
improved on the time-moments of that work, 
we provide an update here. We obtain 
\begin{equation}
a_{\mu}^b = 0.300(15) \times 10^{-10}
\label{eq:amubres}
\end{equation}
This agrees with the value in~\cite{Colquhoun:2014ica} with an 
improvement in uncertainty of a factor of 2.5. 
Since the $b$ quark is so heavy, this is not a significant contribution 
to the anomalous magnetic moment of the muon~\cite{Aoyama:2020ynm}. 

\section{Conclusions}
\label{sec:conclusions}

We have used the fully relativistic HISQ action to calculate the 
masses and decay constants of ground-state bottomonium mesons in lattice QCD including 
the effects of $u$, $d$, $s$ and $c$ quarks in the sea. 
We have used very fine lattices and a range of heavy quark masses 
at each lattice spacing to control the discretisation effects as 
a function of heavy quark mass along with the physical dependence on the 
heavy quark mass of the quantities being studied. 
We have used a fit function with completely generic dependence on the 
heavy quark mass in each of its component pieces, capturing this 
dependence through cubic spline functions.  
Values for bottomonium are obtained by 
evaluating the fit function at zero lattice spacing with tuned sea 
quark masses and a valence quark mass 
tuned to that of the $b$, defined to be the point at the which the $\Upsilon$ mass 
agrees with experiment.
We have also included an analysis of the impact of the electric charge 
of the valence $b$ quarks on the quantities being studied. 
The results given are from the QCD+QED fit but, in all cases, we find 
the impact of QED to be negligible at the level of our uncertainties. 

Our results yield the most precise, to date, lattice 
calculation of the bottomonium hyperfine splitting. We 
obtain the value (repeating Eq.~\eqref{eq:hypincdisc}):
\begin{equation}
\label{eq:hypincdisc2}
M_{\Upsilon} - M_{\eta_b} = 57.5(2.3)(1.0)\ \mathrm{MeV} .
\end{equation}
The first uncertainty is from our fit results (see error budget 
in Table~\ref{tab:errorbudget}) and the second uncertainty is 
from an estimate of missing quark-line disconnected contributions that 
would affect the mass of the $\eta_b$ meson.  
Our result is in agreement with, but on the low side of, 
the experimental average value~\cite{Zyla:2020zbs}. It tends  
to favour the most recent experimental result obtained by the BELLE collaboration 
\cite{Mizuk:2012pb}, although uncertainties (both ours and from the 
experiment) are still too 
large to draw strong conclusions from this.

We also provide the most precise lattice QCD determination of 
the $\Upsilon$ decay constant, which can be used to determine the 
$\Upsilon$ leptonic width. 
Our uncertainty of 1.5\% is three times better than the previous lattice 
QCD calculation of~\cite{Colquhoun:2014ica}. The big advantage of using a 
relativistic formalism, as we do here, is that the vector current can be 
normalised very accurately and nonperturbatively~\cite{Hatton:2019gha}. 
Our result (repeating Eq.~\eqref{eq:fvres}) is 
\begin{equation}
\label{eq:fvres2}
f_{\Upsilon} = 677.2(9.7)\ \mathrm{MeV} ,
\end{equation}
with error budget in Table~\ref{tab:errorbudget}. 
Using this result to obtain the $\Upsilon$ leptonic width gives 
(repeating Eq.~\eqref{eq:ourgamma}): 
\begin{equation}
\label{eq:ourgamma2} 
\Gamma(\Upsilon \rightarrow e^+e^-) =  1.292(37)(3)\ \mathrm{keV}. 
\end{equation}
The first uncertainty is from our result for $f_{\Upsilon}$ and the 
second from possible $\mathcal{O}(\alpha_{\mathrm{QED}}/\pi)$ 
corrections to the formula connecting decay constant and leptonic 
width (Eq.~\eqref{eq:vdecay}). 
This is to be compared with the current experimental 
average of 1.340(18) keV~\cite{Zyla:2020zbs}. We see that our result is 
in good agreement with experiment and our uncertainty 
is just twice as large.

The decay constant of the $\eta_b$ can also be accurately calculated 
with our approach. There is no experimental decay rate that can be 
directly compared to this determination, but the value of $f_{\eta_b}$
is important for our phenomenological understanding of the relationships 
between decay constants for different mesons. We obtain 
(repeating Eq.~\eqref{eq:fpsres})
\begin{equation}
f_{\eta_b} = 724(12)\ \mathrm{MeV} .
\label{eq:fpsres2}
\end{equation}
In particular, repeating Eq.~\eqref{eq:frat}, we find that 
\begin{equation}
\frac{f_{\Upsilon}}{f_{\eta_b}} = 0.9454(99), 
\label{eq:frat2}
\end{equation}
i.e.\ less than 1. This is in contrast to the charmonium 
case where $f_{J/\psi}/f_{\eta_c}$ is larger than 1~\cite{Hatton:2020qhk}. 
Fig.~\ref{fig:fVdivfP} shows how the ratio of the decay constants for vector 
and pseudoscalar heavyonium mesons varies with heavy quark mass. This is qualitatively 
similar to the behaviour seen for the decay constants of 
heavy-light mesons~\cite{Colquhoun:2015oha}.  
Finally, in Fig.~\eqref{fig:Mdivf} we plot the ratios of mass to decay constant 
for pseudoscalar and vector mesons as a function of the ratio of pseudoscalar 
to vector meson masses. These may provide useful information to constrain these 
ratios in QCD-like beyond the Standard Model scenarios. 

The low time moments of the bottomonium vector current-current 
correlator provide a further opportunity to compare lattice QCD 
results to experiment, where the matching inverse-$s$ moments of 
the $b$-quark contribution to 
$R(e^+e^- \rightarrow {\mathrm{hadrons}})$
can be determined. 
Our results for the 4\textsuperscript{th}, 
6\textsuperscript{th}, 8\textsuperscript{th} and 
10\textsuperscript{th} time moments are given in Table~\ref{tab:momfinal} 
where they can be compared to the results obtained from experiment. 
Our uncertainties are 2\% so provide the most accurate test 
to date for these quantities. 
The time moments can be used to determine the $b$ quark contribution to 
the anomalous magnetic moment of the muon. We find (repeating Eq.~\eqref{eq:amubres})
\begin{equation}
a_{\mu}^b = 0.300(15) \times 10^{-10} . 
\label{eq:amubres2}
\end{equation}

Together these results demonstrate how
the properties of low-lying bottomonium states 
can be determined in a fully relativistic calculation in lattice QCD 
and the gains in 
precision that such an approach makes possible.
The results given here also allow us to improve the fully nonperturbative 
determination of the ratio of quark masses, $m_b$ to $m_c$. 
We will present this 
analysis separately. 

\subsection*{\bf{Acknowledgements}} 

We are grateful to the MILC collaboration for the use of
their gluon field configurations and for the use of 
MILC's QCD code. We have modified 
the code to generate quenched U(1) gauge fields and incorporate those 
into the quark propagator calculation as described here.
We are grateful to B. Galloway for contributions to this project 
at a very early stage, and to R. Horgan, C. McNeile and J. Rosner for useful discussions. 
Computing was done on the Darwin supercomputer at the University of
Cambridge High Performance Computing Service as part of the DiRAC facility,
jointly funded by the Science and Technology Facilities Council,
the Large Facilities Capital Fund of BIS and
the Universities of Cambridge and Glasgow.
We are grateful to the Darwin support staff for assistance.
Funding for this work came from the
Science and Technology Facilities Council
and the National Science Foundation.

\begin{appendix}

\begin{table}
\caption{Fit parameters for $F_0$ and $G_0$ for the fit of Eq.~\eqref{eq:spline-fit} to the hyperfine splitting as a function of the vector heavyonium 
mass, $M_{\phi_h}$. The dimensionful constant, $A$ is 0.1 GeV in this case 
and $F_0 = c_F^{(0)} + c_F^{(1)}\times 3\,{\mathrm{GeV}}/M_{\phi_h}$.  
The mean and standard deviation for $c_F^{(0)}$ and 
$c_F^{(1)}$ and the values at the 3 knot positions for $G_0$ are: 
$c_F^{(0)}=0.4407(6371)$; $c_F^{(1)}=0.5031(7476)$; $G_0^{k1}=0.3790(7089)$; 
$G_0^{k2}=0.0956(5041)$; $G_0^{k3}=-0.0338(5285)$. 
The correlation matrix for these 5 parameters is given below. 
These results enable the red fit curve of Figure~\ref{fig:hyp} 
to be reconstructed within the errors given.
}
\begin{tabular}{lllll}
\hline \hline
$c_F^{(0)}$ & $c_F^{(1)}$ & $G_0^{k1}$ & $G_0^{k2}$ & $G_0^{k3}$ \\ 
\hline
1.0 & -0.6195 & -0.1146 & -0.7011 & -0.9420 \\
-0.6195 & 1.0 & -0.7084 & -0.1250 & 0.3225 \\
-0.1146 & -0.7084 & 1.0 & 0.7878 & 0.4386 \\
-0.7011 & -0.1250 & 0.7878 & 1.0 & 0.8980 \\
-0.9420 & 0.3225 & 0.4386 & 0.8980 & 1.0 \\
\hline \hline
\end{tabular}
\label{tab:hyperfine-fit}
\end{table}
\begin{table}
\caption{Fit parameters for $F_0$ and $G_0$ for the 
fit of Eq.~\eqref{eq:spline-fit} to the vector decay constant of the 
vector heavyonium $\phi_h$ meson (upper set) and the decay constant 
of the pseudoscalar heavyonium $\eta_h$ meson (lower set), both 
as a function of the mass $M_{\phi_h}$. 
The dimensionful constant, $A$ is 0.7 GeV in these cases 
and $F_0 = c_F^{(0)} + c_F^{(1)}\times M_{\phi_h}/3\,{\mathrm{GeV}}$.  
The top row of each set 
gives the mean and standard deviation for $c_F^{(0)}$ and 
$c_F^{(1)}$ and the values at the 3 knot positions for $G_0$. 
The correlation matrix for these 5 parameters is given underneath. 
These results enable the red fit curves of both plots 
of Figure~\ref{fig:decay-consts} 
to be reconstructed within the errors given.
}
\begin{tabular}{lllll}
\hline \hline
$f_{\phi_h}$ & & & & \\
\hline 
$c_F^{(0)}$ & $c_F^{(1)}$ & $G_0^{k1}$ & $G_0^{k2}$ & $G_0^{k3}$ \\ 
\hline
0.3487(3783) & 0.1797(110) & 0.030(378) & 0.065(378) & 0.051(378)  \\
\hline
1.0 & -0.0393 & -0.9994 & -0.9987 & -0.9973 \\
-0.0393 & 1.0 & 0.0279 & -0.0080 & -0.0289 \\
-0.9994 & 0.0279 & 1.0 & 0.9986 & 0.9974 \\
-0.9987 & -0.0080 & 0.9986 & 1.0 & 0.9997 \\
-0.9973 & -0.0289 & 0.9974 & 0.9997 & 1.0 \\
\hline \hline
$f_{\eta_h}$ & & & & \\
\hline 
$c_F^{(0)}$ & $c_F^{(1)}$ & $G_0^{k1}$ & $G_0^{k2}$ & $G_0^{k3}$ \\ 
\hline
0.3487(3783) & 0.1797(110) & 0.005(378) & 0.077(378) & 0.122(378)  \\
\hline
1.0 & -0.0399 & -0.9993 & -0.9987 & -0.9962 \\
-0.0393 & 1.0 & 0.0146 & -0.0076 & -0.0372 \\
-0.9993 & 0.0146 & 1.0 & 0.9992 & 0.9977 \\
-0.9987 & -0.0076 & 0.9992 & 1.0 & 0.9992 \\
-0.9962 & -0.0372 & 0.9977 & 0.9992 & 1.0 \\
\hline \hline
\end{tabular}
\label{tab:vecpsdecayfit}
\end{table}
\begin{table}[b]
\caption{Fit parameters for $F_0$ and $G_0$ for the 
fit of Eq.~\eqref{eq:spline-fit} to the ratio of vector to 
pseudoscalar heavyonium decay constants 
as a function of the 
mass of the vector heavonium meson, 
$M_{\phi_h}$. 
$F_0$ is simply a constant, $c_F^{(0)}$ in this case.  
The top row gives the mean and standard deviation for $c_F^{(0)}$ 
and the values at the 3 knot positions for $G_0$. 
The correlation matrix for these 4 parameters is given underneath. 
These results enable the red fit curve of Figure~\ref{fig:fVdivfP} 
to be reconstructed within the errors given.
}
\begin{tabular}{llll}
\hline \hline
$c_F^{(0)}$ & $G_0^{k1}$ & $G_0^{k2}$ & $G_0^{k3}$ \\ 
\hline
0.9973(2080) & 0.0691(2081) & -0.0169(2080) & -0.0538(2082)  \\
\hline
1.0 & -0.9992 & -0.9998 & -0.9988 \\
-0.9992 & 1.0 & 0.9991 & 0.9983 \\
-0.9998 & 0.9991 & 1.0 & 0.9993 \\
-0.9988 & 0.9983 & 0.9993 & 1.0 \\
\hline \hline
\end{tabular}
\label{tab:ratdecayfit}
\end{table}

\section{Reconstructing the heavy quark mass dependence}
\label{sec:appendix}

We give here the fit parameters that enable our fit results for the 
dependence on heavy quark mass of the hyperfine splitting, decay 
constants and ratio of decay constants to be reconstructed. 
The pieces of Eq.~\eqref{eq:spline-fit}
that give the physical curves in the continuum limit are 
$F_0(M_{\phi_h})$ and $G_0(1/M_{\phi_h})$, multiplied by dimensionful 
constant, $A$ (absent for the case of the ratio of decay constants). 
We ignore here the QED pieces of the fit; these have negligible 
effect in all cases. 

$F_0$ is a simple function with 
at most two parameters, $c_F^{(0)}$ and $c_F^{(1)}$. 
$G_0$ is a Steffen spline function~\cite{Steffen:1990} with 3 knots at 2.5, 4.9 and 10.0 GeV in $M_{\phi_h}$, so that in $1/M_{\phi_h}$ they are at
1/2.5, 1/4.9 and 1/10.0.  
Tables~\ref{tab:hyperfine-fit},~\ref{tab:vecpsdecayfit} 
and~\ref{tab:ratdecayfit} 
give the mean and standard deviation of 
$c_F^{(0)}$, $c_F^{(1)}$, and the values at the 3 knots of $G_0$: 
$G_0^{k1}$, $G_0^{k2}$ and $G_0^{k3}$. 
This is followed underneath by the correlation matrix between these 
parameters. The parameters are strongly correlated and this is why we 
give the values to 4 significant figures. 
The splines can easily be implemented using 
the \texttt{gvar} Python 
module~\cite{peter_lepage_2020_4290884}.

\end{appendix} 

\bibliography{heavy}

\end{document}